\documentclass[useAMS,usenatbib,usegraphicx,aas_macros]{mn2e}
\usepackage{amsmath,amssymb,graphicx}
\usepackage[dvips]{color}

\newcommand{\micronn}{\,\hbox{$\mu$m}}

\newcommand\msun{\hbox{\,M$_\odot$}}

\title[Dusty Giants]{The Keck Aperture Masking Experiment: Dust Enshrouded Red Giants}

\author[Blasius et al.]{T. D. Blasius$^{1,2}$, J. D. Monnier$^{1}$, P.G. Tuthill$^3$, W. C. Danchi$^4$, \&
M. Anderson$^{1}$\\
\newauthor 
$^{1}$ Department of Astronomy, University of Michigan, Ann Arbor, MI 48109 USA \\
$^{2}$ California Institute of Technology \\
$^{3}$ University of Sydney, Sydney, Australia \\
$^{4}$ NASA-GSFC, Greenbelt, MD USA 
}

\begin{document}

\date{Submitted in March 2012}


\maketitle

\begin{abstract}
While the importance of dusty asymptotic giant branch (AGB) stars to galactic chemical enrichment is widely recognised, a sophisticated understanding of the dust formation and wind-driving mechanisms has proven elusive due in part to the difficulty in spatially-resolving the dust formation regions themselves. We have observed twenty dust-enshrouded AGB stars as part of the Keck Aperture Masking Experiment, resolving all of them in multiple near-infrared bands between 1.5$\mu$m and 3.1$\mu$m.   
We find 45\% of the targets to show measurable elongations that, when correcting for the greater distances of the targets, would correspond to significantly asymmetric dust shells on par with the well-known cases of IRC~+10216 or CIT~6.   
Using radiative transfer models, 
we find the sublimation temperature of $T_{\rm sub}$(silicates) $=1130\pm90$\,K and $T_{\rm sub}$(amorphous carbon) $=1170\pm60$\,K, both somewhat lower than expected from laboratory measurements and vastly below temperatures inferred from the inner edge of YSO disks.   The fact that O-rich and C-rich dust types showed the same sublimation temperature was surprising as well.
For the most optically-thick shells ($\tau_{2.2\mu{\rm m}}>2$), the  temperature profile of the inner dust shell is observed to change substantially, an effect we suggest could arise when individual dust clumps  become optically-thick at the highest
mass-loss rates.
\end{abstract}




\section{Introduction}
One of the most dramatic phases in the life of an intermediate mass star is the
Asymptotic Giant Branch, a relatively short period where a star loses
most of its initial mass through a dusty wind.  Researchers still do
not understand all the ingredients necessary for producing the high
mass-loss rates observed during this stage.  The massive envelopes
ejected during this phase are thought to be later illuminated during
the planetary nebula stage, a stage where most stars show strong
bipolar circumstellar structures \citep{balick2002}.

Following the advent of infrared detectors, early workers made simple
spherically-symmetric models of dusty shells around large samples of
AGB stars fitting only to the spectral energy
distributions\citep[e.g.,][]{rrh1982,rrh1983,rrh1983b}.  M-type stars are typically surrounded by dust shells composed of amorphous silicates while C-stars have carbonaceous dust.
These early workers were able to
show that dust condensed around 1000~K within a few stellar radii of
the stars and also estimated mass-loss rates typically
10$^{-6}$\msun/yr and as high as 10$^{-4}$\msun/yr.  More recently,
\citet{ivezic1995} developed the code DUSTY to study dust shells in
a systematic way and made models for a large sample of stars, again
fitting just the spectral energy distribution.

The simple picture of spherically-symmetric and uniform mass-loss was
challenged by the observations of the Infrared Spatial Interferometer (ISI), 
a long-baseline mid-infrared interferometer  \citep{danchi1994}. These workers found a
diversity of shell morphologies with some red giants showing episodic dust
shells ejections and others with a more continuous distribution of dust.  
A more dynamic and asymmetric vision of mass-loss fit
into debates into the origins of bipolar symmetry in planetary
nebulae.  High angular resolution near-infrared speckle and aperture
masking on 8-m class telescopes were able to image fine details on
some dust shells, such as the prototype carbon star IRC~+10216
\citep[e.g.][]{tuthill2000}.  An elaborate model was presented by
\citet{menshchikov2002} arguing for complex, spatially-varying dust
properties and density structures.  While IRC~+10216 shows complexity
within the inner few stellar radii, it is unclear if these structures
represent {\em global} asymmetries or just {\em weather conditions} of
the dust formation process observed {\em in situ}.

Here we present the full dataset of dust-enshrouded giants observed
with the 10-year project called the Keck Aperture Masking Experiment
\citep{tuthill2000b}.  This experiment delivered well-calibrated
spatial information on the scale of $\sim$50~milliarcseconds (mas) in the astronomical K band ($\lambda_0=2.2\mu$m), enough
to resolve all the dusty targets presented here and to measure their dust shell sizes and
asymmetries.  This paper includes 20 objects with observations in typically 3
wavelengths ranges, 1.65$\mu$m, 2.2$\mu$m, and 3.1$\mu$m. 
We have also extracted photometry to construct coeval near-IR
spectral energy distributions -- an important factor since these
objects pulsate and show large variations in flux on yearly timescales. 
Lastly, we used a radiative transfer code to fit each epoch of each target star
using simultaneously the NIR photometry and multi-wavelength angular size information from Keck masking.

The primary goals of these observations and modeling efforts are to
measure the physical characteristics of a large sample of the most
extreme dusty AGB stars, 
to address the question of the onset of circumstellar asymmetries, to determine any differences between silicate and carbon-rich dust shells, and to constrain the optical
properties of the dust particles themselves.  Lastly, this publication
marks the final large data release of AGB star data from our
diffraction-limited Keck masking experiment and we anticipate this work
will provide a rich dataset for more detailed modelling efforts by other workers.

\section{Observations}
\subsection{Overview of Observations}
Our observations consist of photometric and visibility data taken on
20 different stars at the W.M. Keck observatory between December 1997
and July 2002.  The wavelengths at which these stars were observed and
the properties of the corresponding filters are listed in Table 1.  A
listing of the observed stars, segregated into carbon-rich and
oxygen-rich groups, along with their basic properties can be found in
Table 2.  Most stars were measured at more than one epoch during this
time span allowing for robust internal data quality checks.

\subsection{Photometric Data} 	
Aperture masking procedures consist of alternating target and
calibrator observations that allow for basic photometry in most
observing conditions.  As part of the standard pipeline
\citep{monnier99, tuthill2000b} we performed aperture photometry on each object, allowing the difference in magnitude ($\Delta$mag) between the target star and calibrator star to be measured. The Vizier catalog service, most often referencing the
Catalogue of Infrared Observations \citep{gezari1999} and 2MASS \citep{2MASS}, was 
used to determine magnitudes at infrared
wavelengths for the calibrators. Interpolation was used
between wavelengths found in the catalogues and the wavelengths at
which our data was taken. 
Occasionally no mid-IR measurements were available for some calibrators and 
we used the calibrator spectral type and the K band flux to estimate the flux density at these longer wavelengths.

As a data quality check we compared our photometry with 2MASS and
found good general agreement, although strict agreement was not expected
since our targets are highly variable and there is some difference in beam sizes. 
We estimated the error on the
photometry points at 10\% based on night-to-night variations.
However, there were instances when we assigned larger errors (between
10 and 32\%) due to saturation 
of the 2MASS photometry used for the calibrator, intrinsic variability of the
calibrator, or effects of cirrus clouds in some of the original data.   Indeed, there were some nights too contaminated by variable clouds to allow photometry to be extracted at all.

Table 3 is a journal of observations, including the observing date(s),
the filter(s) used, the aperture mask(s) used, and calibrator star
name.  We have compiled the adopted calibrator properties in Table 4.

\subsection{Visibility Data}
\subsubsection{Methodology}
Our group carried out aperture masking interferometry at the Keck-1
telescope from 1996 -- 2005.  We have published images and size
measurements with (at the time) unprecedented angular resolution on topics ranging
from young stellar objects, carbon stars, red supergiants, and
photospheric diameters of Mira variables
\citep[e.g.,][]{monnier99a,tuthill2000,tuthill2000b,danchi2001}.

The NIRC camera with  the image magnifier \citep{matthews96} was used in
conjunction with the aperture masking hardware to create fringes at
the image plane.  The data frames were taken in speckle mode ($T_{\rm
  int}$=0.14\,s) to freeze the atmosphere.  In the work presented here, 
multiple aperture masks and bandpass filters were employed.
After flat-fielding,
bad pixel correction, and sky-subtraction, Fourier methods were used
to extract fringe visibilities and closure phases from each frame and
averaged in groups of 100 frames.  Absolute calibration to account for
the optical transfer function and decoherence from atmospheric seeing
was performed by interleaving science observations with measurements
of unresolved calibrators stars.  At the end of the pipeline, the data
products are purely interferometric as if obtained with a
long-baseline interferometer.  A full description of this experiment
can be found in \citet{tuthill2000b} and \citet{monnier99}, with further discussion of
systematic errors in \citet{monnier2004a} and \citet{monnier2007}.
All V$^2$ and closure phase data are available
from the authors; all data products are stored in the FITS-based,
optical interferometry data exchange format (OI-FITS), as
described in \citet{pauls2005}.

\subsubsection{Basic Results}
Before undertaking radiative transfer modeling, 
we provide the results of basic geometrical 
analysis of the visibility data.  The simplest representation of the data is generally a circularly-symmetric Gaussian 
envelope, a useful model to give a characteristic size to the emission.  
Table 5 provides the visibility intercept ($V_0$) and the Full-width at Half-maximum (FWHM) for the best fit for all datasets, including the reduced {$\chi^2$}.
Errors are generally dominated by systematics related to the calibration procedure (i.e., seeing variation between source and calibrator visits) and we have used the relations established in
\citet{monnier2007} to quantify our errors. In some cases, there was evidence of two components to the visibility curve and we have also fitted a slightly more complex model of a point source plus a Gaussian envelope to all epochs. Table~6 contains the best fit parameters of the 2-component model,  including the estimated fraction of light in the point source ($f_{\rm point}$) and the fraction of light in the Gaussian envelope ($f_{\rm Gauss}$).

In addition,we fitted each object with a 2-dimensional Gaussian function in order to search for signs of asymmetry.  Objects with observed asymmetry are marked with an
asterisk in Table~5 .  Table 7 lists all the object with confirmed asymmetries and we include the amount of elongation 
($\frac{{\rm FWHM}_{\rm major}}{{\rm FWHM}_{\rm minor}}$) and the position angle (degrees East of North) of the major axis.  Here we have used the spread of  measured position angles between wavelength channels and epochs  to estimate the PA error. We will discuss further these findings in \S\ref{discussion}.

\section{Dust shell Modeling}
\subsection{Introduction}
The objects in our study all have spectral energy distributions that peak in the infrared.
Indeed, these stars are surrounded by dust shells that
absorb the stellar light and then reemit the energy in the infrared.  
In order to extract physical characteristics of these dust shells (i.e., optical depths, temperatures, etc),
we must be able to compute how the dust will
absorb, scatter, and reemit the energy from the star.  We accomplish
this with the radiative-transfer model
DUSTY \citep{Ivezic1999}.  While DUSTY is limited to calculations in spherical symmetry, 
we established in the previous section that most of our objects show only mild signs of global asymmetries; however, we caution that our results will be suspect for the most asymmetric of the targets listed in Table~7.
Given a small number of input parameters, DUSTY can quickly compute synthetic photometry and intensity profiles for dust shells. These outputs can then be compared to
the data that we have experimentally obtained.

\subsection{Model Description}
We applied a uniform procedure for fitting all of our objects.  Here we discuss which properties were held fixed and how we explored a grid of the key dust shell parameters.

We begin with the central star. At the beginning of our study we used a featureless Planck blackbody spectrum, however we came
to realise that a blackbody spectrum is a rather poor approximation for the extremely late-type giants in our sample due to strong molecular absorption bands.
Most notably, the HCN absorption
feature of carbon-rich stars sits directly at the PAHcs (3.0825~$\mu$m) wavelength, where we have many observations.   
Because of the severe optical absorption of the dust, spectral types are not known for most stars in our sample and we have adopted an effective temperature of 2600K for all stars, which is as cool as we could find converged synthetic spectra.  For the carbon stars we used a MARCS model as described
in \citet{loidl2001} and for the M-giants we used a PHOENIX NEXTGEN model as described in
\citet{hauschildt1999}. The medium-resolution synthetic spectra from these sources were smoothed before input into DUSTY.  Unfortunately, we do not have useful distance estimate to our sources -- so we adopted a distance of 1000~pc and interstellar reddening of $E_{B-V}=0.5$ for all objects.
 We note that the dust shells around the stars absorb nearly all of the energy from the central source, acting as a kind of calorimeter.  Thus, while our 2600K estimate for the central star temperature is crude, we expect the bolometric luminosity (for assumed $d=1000~$pc) to be more accurate. However in practice our luminosity estimates are poor due to uncertainties in the dust shell optical depth and the fact we are not integrating the whole observed SED throughout the mid- and far-infrared.
   
Based on the shape of the SED (and the presence of a silicate feature in IRAS-LRS spectra), we determined each star to have either carbon-rich dust or silicate-rich dust.  Based on this assignment, we chose amorphous carbon \citep{hanner1988} or warm amorphous silicates \citep{ossenkopf1992} respectively in the DUSTY model setup.
\cite{speck2008} discussed how silicates close to AGB stars could quickly anneal to crystalline grains but a full exploration of optical constants for different grain types was beyond the scope of this work.  For the grain size distribution, we adopted the standard MRN power-law grain size distribution between  0.005-0.25 $\mu$m \citep{Mathis1977}; a later exploration of larger grain sizes did not systematically improve fits \citep[also see discussion by][]{speck2009}.    
Another property of the dust shell we fixed is that the dust density follows a $r^{-2}$ power-law, corresponding to constant mass-loss rate. 

Lastly, we come to the parameters of the model that are not fixed:
the temperature of the dust shell at the inner boundary, T$_{\rm dust}$, the radius of
the star, R$_{\rm star}$, and the K-band optical depth $\tau_{2.2\mu\,m}$ of the dust shell 
(as integrated along the line-of-sight from the observer to the star). 
In the next section, we explain our fitting procedure.

\subsection{Fitting Methodology}
We explored inner dust temperatures T$_{\rm dust}$ between 400K--1500K.  This
range explored both the high temperatures thought to be
prohibitive of dust creation and low temperatures too cool for steady-state
dust production.  Note that when setting up a model in DUSTY, one does not specify the inner radius of the dust shell: this quantity is calculated based on the luminosity of the star and the specific inner shell dust temperature T$_{\rm dust}$.  In terms of optical depth, we explored  $\tau_{2.2\mu\,m}$
between 0 and 9.  This range provided a full fitting region for our
objects and values of $\tau_{2.2\mu\,m}$ much above 9 were too computationally
expensive.  Finally, R$_{star}$ was recognised to simply be a scaling
factor for the model outputs and could easily be optimised for every pair of ($T_{\rm dust}$, $\tau_{2.2\mu\,m}$).  Because the DUSTY calculation was fast and we only had to
optimise over a few parameters, we chose to carry out an exhaustive grid calculation over all ($T_{\rm dust}$, $\tau_{2.2\mu\,m}$).

For each location in the grid we calculated the model SED as well as the radial intensity profiles.  We calculated a $\chi^2$ based on both our coeval near-infrared photometry as well as Keck masking visibility curves. For the SED, we also used including V-band magnitudes in our fit with a very low weight to ensure that the optical depths were not too low (important especially when for objects without photometry in all three near-IR wavelength bands).  When calculating the $\chi^2$ for the visibility curves, we adopted the following procedure.
Because the y-intercept of our observed visibility data can fluctuate $\pm$5\% due to seeing calibration, we normalised each visibility to 1.0 at zero baseline before fitting.  
Also, we weighted the visibility points so that the SED and the visibility data were separately given equal weight in the final reduced $\chi^2$. We purposefully chose not to include longer wavelength SED measurements, such as IRAS data, in our fitting. 
By fitting only to near-IR photometry and near-IR spatial data we can isolate and only probe dust emitted within the last few decades. This allows us to keep the model as simple as possible and enhances  the validity of our assumption of constant mass loss rate (ie., $\rho\propto r^{-2}$).

Once the grid calculation over inner dust temperature T$_{dust}$ and $\tau_{2.2\mu\,m}$ was completed, the $\chi^{2}$ surface was used to estimate the best-fit parameters.  The uncertainty estimates were produced by considering the region where the reduced $\chi^2$ was less than 2, a {\em highly conservative criterion} that reflects the highly-correlated errors in our datasets.  In cases where the best-fit $\chi^2$ is above 1, we scaled the $\chi^2$ results by the best-fitting value before estimating the parameter uncertainties. 
The best-fitting parameters and their uncertainties are compiled in Table~8.

In addition to providing the fitting results in tabulated form, 
we also include here a series of figures which graphically represent the
new data, modeling results, and the $\chi^2$ surface in our grid. 
These plots can be found in each of
Figures 1--20.  The first panel in each figure contains the observed near-IR photometry and best-fit model SED. The second
panel in each figure contains the multi-wavelength visibility curves averaged azimuthally along with the model curves.  Finally, the third panel shows the $\chi^2$ surface in the ($T_{\rm dust}$, $\tau_{2.2\mu\,m}$) plane.
We have grouped all the epochs for the same object together so one can see the self-consistency in the derived dust shell parameters -- indeed, consistent dust shell properties were recovered when fitting to different epochs, despite large changes in the central star luminosity due to pulsations.

One of the most important results to take away from these panels that we clearly break
the standard degeneracy between dust temperature and optical depth. This is because of our new spatial information -- by measuring the {\em sizes} of the dust shell at various wavelengths we can simultaneously constrain the temperature and optical depth.  In the past, one typically had to choose an inner dust temperature based on physical arguments concerning the dust condensation temperatures of various dust species. Here, we see that the inner dust temperature can be constrained independently from other parameters and the implications are discussed further in the next section.

While the simultaneous fits to the near-IR SED and visibility data were generally acceptable, we found the fits to the shortest wavelength visibility data at H band were systematically worse.   Since this band is most sensitive to scattering by dust, we explored modified dust distributions, especially using larger grains; we did not find systematic improvements to the fits by altering dust size distribution from MRN or by using other dust constants.  

\section{Discussion}
\label{discussion}
Our survey provides the first constraints on the asymmetry of the dust shells for such a large sample of  dust-enshrouded AGB stars.
We found that 4 out of 7 M-stars and 5 of 13 C-stars showed evidence of dust shell asymmetries, with dust shell elongations between 10\% and 40\%. While this level of asymmetry may sound mild, it actually (quantitatively) compares to the level of asymmetry that would be expected for the most asymmetric dust shells known if placed at 1~kpc.  For instance, we know that IRC~+10216 \citep{tuthill2000} and CIT~6 \citep{monnier2000} have dramatic global asymmetries in their dust shell, detailed imaging made possible by virtue of their proximity. If we placed these targets farther away, we would not be able to image the detail but they would appear $\sim$20\% elongated, similar to the degree observed here in 45\% of our sample.  For CIT~3, we confirm the asymmetries seen by \citet{hofmann2001} and note that  \citet{vinkovic2004} showed that the 20\% elongation could be explained by a bipolar outflow. 
That said, 
clumpy dust formation \citep{fleischer1992} might also cause stochastic variations in the inner dust shell geometry that could appear as short-lived elongations.   Mid-infrared observations with long-baseline interferometers (e.g, ISI, VLTI-MIDI) should focus on these targets to determine the nature of the asymmetries.  In addition, long-term monitoring of these dust shells will help settle debates concerning when the environments of evolved stars develop large scale asymmetries commonly revealed in the later planetary nebula stage.  For instance, a long-term asymmetry in a constant position angle (as judged by linear polarization or spatially resolved data) would be a sign of a global bipolar mass-loss asymmetry and not just {\em weather}.
	
In order to look at dust shell properties for our full sample, we have plotted the inner edge dust temperature
$T_{\rm dust}$ vs total dust shell optical depth $\tau_{2.2\mu\,m}$ for all our targets.  Figure 21 shows these results split into O-rich and C-rich dust type.   For K-band optical depths below 2, we find the sublimation 
temperature of $T_{\rm sub}$(silicates) $=1130\pm90$\,K and $T_{\rm sub}$(amorphous carbon) $=1170\pm60$\,K, both somewhat lower than expected from laboratory measurements \citep{lodders1999} and vastly below temperatures inferred from the inner edge of YSO disks \citep[$\sim$1800K,][]{tannirkulam2008, benisty2010}.   
One component to the observed lower dust temperature could be due to the fact that the central star varies in luminosity by about a factor of 2 during the pulsation cycle and we see the dust cooler than the condensation temperature during phases away from maximum light.

The $T_{\rm dust}$ vs optical depth $\tau_{2.2\mu\,m}$ diagram (Figure~21) also
shows  no statistically-significant difference between O-rich and C-rich dust types,  counter to expectation of higher temperatures for carbon-rich dust  \citep{lodders1999}.   We recognise that our simple dust shell modeling may not lead to accurate estimates of the
dust sublimation temperature if the inner dust formation environment radically departs from a power law density distribution, 
perhaps due to pulsations, timescale for dust formation, or multiple dust species.  Interestingly though these concerns would likely affect C-rich and O-rich shells similarly and so the lack of a clear difference in sublimation temperatures between these dust types appears robust.

	
The other important feature of Figure 21, $T_{\rm dust}$ vs optical depth $\tau_{2.2\mu\,m}$, is the
apparent temperature at the inner edge of the dust shell gets lower and lower with increasing optical depths above 2.  This appears true for both C-rich and O-rich shells.  Here we do not believe we are seeing an actual reduction in the dust sublimation temperature, but
rather a change in the temperature profile in the inner dust formation zone due to a breakdown in the assumption of a spherically-symmetric $r^{-2}$ density power law.
We have ample evidence that dust formation is clumpy, as has been 
imaged in great detail for IRC +10216 \citep{tuthill2000}, but these clumps have been shown to have a relatively weak affect on the temperature structure for low optical depths.  Next we further explore how a clumpy dusty environment could change the temperature profile of the dust shell when the individual clumps become themselves optically thick to the stellar and even hot dust radiation field.

Clumpy structures are seen to evolve in 2-D models
of dust shells due to self-amplifying density perturbations
\citep[e.g.][]{Woitke2000}.   First
optically thick dust regions form and these regions cast shadows on
the dust behind them.  Consequently, the temperatures decrease by
100's of degrees K and this allows for a higher rate of dust formation
in these shadow regions.  Scattering and re-emission of light by the
optical thick regions increases the intensity of radiation between
them and eventually the light escapes through the optically thin
regions in between the optically thick regions.  Thereupon, the
temperature within the optically thin regions increases, which
decreases the rate of dust production.  These processes thus amplify
the initial homogeneities until large-scale clumpy structures start to
form, such as ``dust fingers" \citep{Woitke2005}.  Indeed, \citet{Woitke2005} did see {\em average dust temperatures to be reduced} due to these opacity effects but at much weaker level than 
we see in Figure~21.  Realizing that our data reveal a strong effect only at $\tau$'s several times larger than probed by \citet{Woitke2005}, we suggest that dust shadowing effects get dramatically stronger  when individual clumps become optically thick to both stellar radiation as well as hot dust emission. 
A 3D radiative transfer calculation of a dusty dust shell could validate or disprove this explanation.

In conclusion, our large sample of  spatially resolved dust-enshrouded stars have led to new insights into the late stages of AGB star evolution. We find levels of dust shell elongations that point to significant asymmetries in nearly half of our targets.  Our spatial and SED data combined has eliminated some model degeneracies, and we now have the best constraints on the actual sublimation temperatures for dust forming in this outflows, finding lower temperatures than expected from terrestrial experiments and not confirming the large difference expected between carbon-rich and silicate-rich dust. Lastly, we discovered a systematic change in the temperature profile for inner-most dust regions when the dust shell optical depth rises above $\tau_{2.2\mu\,m}>2$.  This observed lowering of the central dust temperatures could be naturally explained as a consequence of shadowing caused by clumpy dust formation on spatial scales smaller than our angular resolution, but other possibilities should be further explored as well.

\section*{Acknowledgements}
We thank Dr. Charles Townes for his long-standing support of this work.  We thank Angela Speck for her insightful comments upon reading a draft of this manuscript.  We also acknowledge interesting discussions with Peter Woitke regarding the effect of
clumpy structures on the temperature profile, and we thank Rita Loidl Gautschy for her help in acquiring the C-star synthetic spectrum.  This research
has made use of the SIMBAD database, operated at CDS, Strasbourg,
France. This publication makes use of data products from the Two
Micron All Sky Survey (2MASS), which is a joint project of the
University of Massachusetts and the Infrared Processing and Analysis
Center/California Institute of Technology, funded by the National
Aeronautics and Space Administration and the National Science
Foundation.  The data presented herein were obtained at the W.M. Keck
Observatory, which is operated as a scientific partnership among the
California Institute of Technology, the University of California and
the National Aeronautics and Space Administration.  The Keck
Observatory was made possible by the generous financial support of the
W.M. Keck Foundation.  The authors wish to recognise and acknowledge
the very significant cultural role and reverence that the summit of
Mauna Kea has always had within the indigenous Hawaiian community.  We
are most fortunate to have the opportunity to conduct observations
from this mountain.

\bibliographystyle{apj}
\bibliography{Monnier_refs,WRSizes,Review}

\begin{table*}
\centering
\caption{\label{filters} Properties of NIRC Camera Infrared Filters. Reference: The NIRC Manual.} 
\begin{tabular}{llll}
Name & Center Wavelength & Bandpass FWHM & Fractional \\
& $\lambda_0$ (\micronn) & $\Delta\lambda$ (\micronn) & Bandwidth \\
\hline
FeII & 1.6471 & 0.0176 & 1.1\% \\
H & 1.6575 & 0.333 & 20\% \\
K & 2.2135 & 0.427 & 19\% \\
Kcont & 2.25965 & 0.0531 & 2.3\% \\
CH4 & 2.269 & 0.155 & 6.8\% \\
PAHcs & 3.0825 & 0.1007 &3.3\% \\
\hline
\end{tabular}
\end{table*}

\begin{table*}
\caption{Basic Properties of Targets\label{targets}}
\begin{tabular}{lllllllllll}
Source & RA (J2000) & Dec (J2000) &
V & J\textsuperscript{a} & H\textsuperscript{a} & K\textsuperscript{a} &
Spectral \\
Names & & & mag & mag & mag & mag & Type\\
\hline

AFGL 230  & 01 33 51.21 & $+$62 26 53.5 & --- & 16.747 & 11.232 & 7.097 & M$^{(7)}$ \\
AFGL 2019 & 17 53 18.9 & $-$26 56 37 & 20.2$^{(2)}$  & 6.338 & 4.035 & 2.616 & M8$^{(1)}$ \\
AFGL 2199  & 18 35 46.48 & $+$05 35 46.5 & --- & 8.04 & 4.85 & 2.701 & M$^{(6)}$\\
AFGL 2290 & 18 58 30.02 & $+$06 42 57.7 & --- & 13.169 & 8.966 & 5.862 & M$^{(6)}$ \\
CIT 1 & 00 06 52.94 & $+$43 05 00.0 &  9.00$^{(1)}$ & 3.041 & 1.829 & 1.115 & M9$^{(1)}$ \\
CIT 3 & 01 06 25.98 & $+$12 35 53.0 & ---- &  7.45 & 4.641 & 2.217 & M9$^{(1)}$ \\
v1300 Aql & 20 10 27.87 & $-$06 16 13.6 & 20$^{(1)}$ & 6.906 & 3.923 & 2.059 & M$^{(1)}$ \\
\hline
AFGL 1922 & 17 07 58.24 & $-$24 44 31.1 & --- & 12.244 & 9.181 & 6.342 & C$^{(3)}$ \\
AFGL 1977 & 17 31 54.98 & $+$17 45 19.7 & 9.9$^{(4)}$ & 10.536 & 7.994 & 5.607 & C$^{(1)}$ \\
AFGL 2135 & 18 22 34.50 & $+$27 06 30.2 & --- & 9.043 & 6.002 & 3.643 & C$^{(1)}$ \\
AFGL 2232 & 18 41 54.39 & $+$17 41 08.5 & 9.7$^{(1)}$ & 5.742 & 3.444 & 1.744 &  C$^{(1)}$ \\
AFGL 2513 & 20 09 14.22& $+$31 25 44.0 & --- & 8.229 & 5.705 & 3.69 & C$^{(1)}$ \\
AFGL 2686 & 20 59 08.88 & $+$27 26 41.7 & 20$^{(1)}$ & 9.112 & 6.268 & 4.075 & Ce$^{(1)}$ \\
AFGL 4211 & 15 11 41.89 & $-$48 20 01.3 & --- & 10.711 & 7.751 & 5.154 & C$^{(3)}$  \\
IRAS~15148-4940 & 15 18 22.05 & $-$49 51 04.6 & 11.8$^{(1)}$ & 5.297 & 3.071 & 1.696 & C$^{(1)}$ \\
IY Hya & 10 17 00.52 & $-$14 39 31.4 & 14$^{(1)}$ & 5.919 & 3.666 & 1.964 & C$^{(5)}$ \\
LP And & 23 34 27.66  & $+$43 33 02.4  &  ---  & 9.623 & 6.355 & 3.859 & C$^{(1)}$ \\
RV Aqr & 21 05 51.68 & $-$00 12 40.3 &11.5$^{(1)}$ & 4.046 & 2.355 & 1.239 & C$^{(5)}$ \\
v1899 Cyg & 21 04 14.8 & $+$53 21 03 &15.6$^{(1)}$ & 10.84 & 8.693 & 6.596 & C8$^{(5)}$ \\
V Cyg & 20 41 18.2702 & $+$48 08 28.835 & 7.7$^{(1)}$ & 3.096 & 1.273 & 0.117 & C$^{(1)}$ \\
\hline
\end{tabular}
\begin{flushleft}
\textsuperscript{a} ~ These
  magnitudes (from 2MASS) are merely
  representative since the targets are variable. See Table 3 for our new photometry.
\newline \textbf{Note:} Horizontal line separates oxygen-rich (top) from carbon-rich (bottom). \newline \textbf{References:} (1) Simbad,
(2) \citet{Monet1998}, (3) \citet{Buscombe1998}, (4) \citet{Egret1992},
(5) \citet{Skiff2007}, (6) \citet{Volk2002}  (7) \citet{Garcia-Hernandez2007}
\end{flushleft}
\end{table*}

\begin{table*}
\tiny
\caption{Journal of observations and derived photometry\label{observations}}
\begin{tabular}{llllll}
Target & Date(s) & Filter & Aperture &Magnitude &Calibrator\\
 & (UT) &    & Mask & & Names\\
\hline
AFGL 230   & 1997 Dec & k          & FFA  & 8.34$\pm$0.1 & $\chi$~Cas\\
                  &                    & PAHcs  & KL Relation*  & 5.11$\pm$0.2 & \\
                  & 2002 Jul    & k         & FFA              & 8.99$\pm$0.1 & HD 9878\\
                  &		    & PAHcs & FFA             & 5.90$\pm$0.3 & HD 9329\\
AFGL 2019 & 2000 Jun   & CH4     & annulus 36 & 2.48$\pm$0.1 & HD 163428\\
                   &		     & h         & annulus 36 & 3.84$\pm$ 0.1 & HD 156992\\
                   &                    & PAHcs & annulus 36  & 1.52$\pm$0.1 & HD 163428\\
AFGL 2199 & 1998 Apr & CH4     & annulus 36  & 2.99$\pm$0.1 & HD 170137\\
                  &                   & PAHcs & annulus 36  & 1.80$\pm$0.1 & HD 170137\\
AFGL 2290 & 1998 Jun      & CH4         &  annulus 36  & 4.72$\pm$0.1 &HD 173074\\ 
                   &                       & PAHcs      &annulus 36  & 2.60$\pm$0.1 &HD 173074\\ 
                   & 1999 Apr     & CH4          &annulus 36  & 5.61$\pm$0.1 &HD 173833\\ 
                   &                       & k               &annulus 36  & 6.19$\pm$0.32 & HD 231437\\
                   &                       & PAHcs       &annulus 36  & 3.29$\pm$0.1 &HD 173833\\  
CIT 1        & 2000 Jun & CH4  & annulus 36 & 2.60$\pm$0.1 & $\lambda$~And \\
                &                    & h       & annulus 36 & 4.18$\pm$0.25 & HD 222499\\
                &			& PAHcs & annulus 36 &1.51$\pm$0.1 & $\lambda$~And\\
CIT 3        & 1997 Dec & Kcont  & annulus 36 & 1.08$\pm$0.1 & $\delta$~Psc \\
	       &                    & PAHcs  & annulus 36 & -0.14$\pm$0.1 & $\delta$~Psc\\
	       & 1998 Sep & CH4     & Golay 21      & 2.45$\pm$0.1 & $\delta$~Psc\\ 
	       &                    & PAHcs & Golay 21     & 1.04$\pm$0.1 & $\delta$~Psc\\
v1300 Aql & 1998 Jun    & CH4        & annulus 36   & 1.39$\pm$0.1 & HD 189114\\
                   &                      & h             & annulus 36  & 3.29$\pm$0.25 & HD 192464\\
                   &                      & PAHcs     &annulus 36  & 0.60$\pm$0.1 &HD 189114\\
                   & 1999 Jul      & kcont      & annulus 36  & 2.02$\pm$0.1 & SAO 14382\\
                   &                       & PAHcs     & annulus 36  & 0.86$\pm$0.1 & SAO 14382\\
\hline
AFGL 1922 & 2000 Jun   & k          & annulus 36  & 6.34$\pm$0.25 & HD 156992\\
                   &                     & PAHcs & KL relation* & 3.62$\pm$0.25 &      \\
                   & 2001 Jun   & k         & annulus 36   & 4.90$\pm$0.1 & HD 158774\\
                   &                     & PAHcs & KL relation*  & 2.32$\pm$0.25 &   \\
AFGL 1977 & 1998 Jun      & CH4         &annulus 36  & 4.19$\pm$0.1 &  HD 158227\\
                  &                        & h             &annulus 36  & 7.05$\pm$0.1 & HD 158227\\   
                  &                        & PAHcs     &annulus 36  & 1.84$\pm$0.1 & HD 157049\\
                  & 1999 Apr      & CH4          &annulus 36  & 2.77$\pm$0.1 & HD 157049\\
                  &                        & PAHcs     & annulus 36  & 0.59$\pm$0.1 & HD 157049\\
AFGL 2135   & 2001 Jun     & k              &  annulus 36 & 3.29$\pm$0.1 & HD 168366, HD 181700\\
                     &                     & PAHcs     &  annulus 36 & 1.27$\pm$0.3 & HD 177716\\
AFGL 2232  & 1998 Jun     & CH4           &annulus 36  & 2.04$\pm$0.1 &  HD 158227\\
                    &                      & h                &annulus 36  & 4.12$\pm$0.1 & HD 158227\\
                    &                       &PAHcs       &annulus 36  & 0.68$\pm$0.1 & HD 157049\\
                    &                      & CH4          & Golay 21      & 2.28$\pm$0.3 & HD 168720\\
                    &                       & PAHcs     &Golay 21      & 0.94$\pm$0.3 & HD 168720\\
                    & 1999 Apr     & CH4          & annulus 36     & 1.06$\pm$0.1 & HD 173833\\
                    &                       & PAHcs      & annulus 36    & -0.38$\pm$0.1 & HD 173833\\
AFGL 2513 & 1998 Sep  & h          & annulus 36  & 6.58$\pm$0.1 & HD 196241\\ 
                  &                     & CH4     & annulus 36  & 4.03$\pm$0.1 & HD 200451\\
                   &                    & PAHcs & annulus 36  & 3.16$\pm$0.3 & $\epsilon$~Cyg\\
                  & 1999 Jul    & CH4      & annulus 36  & 2.90$\pm$0.1 & HD 188947\\
                  &                     & PAHcs  & annulus 36  & 1.70$\pm$0.1 & HD 188947\\
AFGL 2686 & 1998 Sep  & CH4       & annulus 36  & 2.95$\pm$0.1 & HD 200451\\
                  &                     & h           & annulus 36  & 5.82$\pm$0.21 & HD 200451\\
                  &                     & PAHcs   & annulus 36 & 1.01$\pm$0.1 & $\epsilon$~Cyg, $\lambda$~And\\
                  & 1999 Jul     & CH4       & annulus 36  & 5.13$\pm$0.1 & HD 188947\\
                  &                     & PAHcs    &  annulus 36  & 2.92$\pm$0.1 & HD 188947\\
                  &                      & h           &  annulus 36  & 8.48$\pm$0.3 & HD 198330\\
AFGL 4211  & 2000 Jun  & CH4     & annulus 36  & 3.62$\pm$0.3 & HD 137709\\
                   &                    & PAHcs & annulus 36  & 1.42$\pm$0.3 &HD 137709\\
                   & 2001 Jun  & k           & annulus 36 & 4.70$\pm$0.1 & HD 137709\\
                   &                     & PAHcs & KL relation* & 2.64$\pm$0.2 &  \\
IRAS~15148-4940 & 2001 Jun & CH4   & annulus 36 & 1.25$\pm$0.3 & HD 137709\\    
                               &                    &k         &  annulus 36 & 1.30$\pm$0.3 & HD 137709\\ 
                              &                     & PAHcs &  annulus 36 & 1.71$\pm$0.1 & HD 136422\\ 
IY Hya          & 1999 Apr      & CH4         & annulus 36  & 2.08$\pm$0.1 & HD 87262\\
                     &                       & PAHcs     & annulus 36  & 1.37$\pm$0.1 & $\mu$~Hya\\
LP And     & 1998 Sep   & CH4       &  annulus 36  & 3.89$\pm$0.1 & HD 222499, $\lambda$~And\\
                  &                     & h            & annulus 36  & 7.05$\pm$0.25 &HD 222499\\
                  &                     & PAHcs    &  annulus 36  & 1.72$\pm$0.1 & $\lambda$~And\\
                  & 1999 Jul     & CH4        & Golay 21        & 4.01$\pm$0.1 & $\alpha$~Cas\\
                  &                      & PAHcs    & Golay 21        & 1.80$\pm$0.1 & $\alpha$~Cas\\
                  & 1999 Jan    & CH4        & Golay 21        & 3.26$\pm$0.1 & $\alpha$~Cas\\
                  &                      & PAHcs    & Golay 21        & 1.18$\pm$0.1 & $\alpha$~Cas\\
RV Aqr           & 1999 Jul      & CH4        & Golay 21 & 1.23$\pm$0.25 & SAO 143482, 3~Aqr\\ 
                     &                       & PAHcs     & Golay 21 & 0.56$\pm$0.25 & SAO 143482, 3~Aqr\\ 
                     & 1998 Jun     & CH4         &  Golay 21 & 1.52$\pm$0.1 & HD 196321\\
                     &                       & PAHcs     &   Golay 21 & 1.15$\pm$0.1 & HD 196321\\
v1899 Cyg & 1998 Jun     & CH4        &  annulus 36 & 5.53$\pm$0.1 & HD 202897\\
                     &                       & h            & annulus 36 & 7.87$\pm$0.3 & HD 200817\\
                     &                       & PAHcs    &  annulus 36 & 3.71$\pm$0.1 & HD 202897\\
                     & 1999 Jul      & k              &  annulus 36 & 6.40$\pm$0.1 & HD 198661\\
                     &                       & PAHcs    & KL relation*   & 4.72$\pm$0.2 &  \\
V Cyg          & 1998 Jun     & feii           &  annulus 36  & 2.59$\pm$0.1 & HD 192909\\
                   &                        & kcont      &  annulus 36  & 0.53$\pm$0.1 & HD 192909\\
                   &                        & CH4         &  Golay 21  & 0.50$\pm$0.1 & HD 192909\\
                   &                        & PAHcs     & annulus 36  &  0.26$\pm$0.1 & HD 192909\\
                   &                       &   PAHcs     & Golay 21     & 0.19$\pm$0.1 & HD 192909\\
                   & 1999 Apr     & CH4           & Golay 21  & 0.15$\pm$0.1 & $\xi$~Cyg\\
                   &                       & PAHcs       & Golay 21  & -0.25$\pm$0.1 & $\xi$~Cyg\\
                   & 2001 Jun     & CH4           & annulus 36 & -0.27$\pm$0.1 & $\xi$~Cyg\\
                   &                       & PAHcs      & annulus 36 & -0.69$\pm$0.1 & $\xi$~Cyg\\
\hline
\end{tabular}
\begin{flushleft}
* ~ This point was extrapolated from another epoch for the same star and assigned an error of 0.2 mag.
\newline 
\textbf{Note:} Horizontal line separates oxygen-rich (top) from carbon-rich (bottom).
\end{flushleft}
\end{table*}

\begin{table*}
\caption{Basic Properties of Calibrators}
\begin{tabular}{lllllllllll}
Calibrator & J & H & K & PAHcs & Reference\\
 & mag & mag& mag & mag & \\
\hline
HD 168720 & 1.79 & 0.875 & 0.870 & 0.794 &  \citet{McWilliam1984}, \citet{2MASS}, \citet{Neugebauer1969}\\
HD 170137 & 3.476 &  2.737 & 2.230 & 2.16 & \citet{2MASS}, \citet{Neugebauer1969}\\
$\epsilon$~Cyg  & 0.641 & 0.2 & 0.1 & 0.011 & \citet{Neugebauer1969}, \citet{Ghosh1984}, \citet{Price1983}\\
HD 200451  & 4.101 & 3.231 & 2.840 & --- & \citet{2MASS}, \citet{Neugebauer1969}\\
HD 231437 & 5.027  & 3.958 & 3.693 & --- & \citet{2MASS}\\
HD 173833 & 3.488 & 2.647 & 2.1 & 2.02 & \citet{2MASS}, \citet{Neugebauer1969}\\
HD 158227 & 5.626 & 4.984 & 4.812 & --- & \citet{2MASS}\\
HD 157049 & 1.975 & 1.149 & .830 & .684 & \citet{2MASS}, \citet{Neugebauer1969}, \citet{Price1983}\\
HD 168366 & 5.049 & 4.535 & 4.255 & --- & \citet{2MASS}\\
HD 181700 & 3.938  & 2.993 & 2.735 & --- & \citet{2MASS}\\
SAO 143482 & 1.665 & 0.790 & 0.573 & 0.436 & \citet{2MASS}, \citet{Gullixson1983}\\
HD 189114 & 3.212 & 2.030 & 1.953 & 1.908 & \citet{2MASS}, \citet{Gosnell1979}\\
HD 137709 & 2.232 & 1.532 & 1.331 & 1.257 & \citet{2MASS}, extrapolation\\
HD 222499 & 4.641 & 3.804 & 3.627 & --- &  \citet{2MASS}\\
$\lambda$~And & 1.970 & 1.4 & 1.287 & 1.245 & \citet{Johnson1966}, \citet{Price1983}, \citet{Selby1988}\\
HD 9878 & 6.631 & 6.730 & 6.698 & --- & \citet{2MASS}\\
HD 9329 & 4.961 & 4.381 & 4.341 & 4.29 & \citet{2MASS}, extrapolation\\
HD 156992 & 3.901 & 3.123 & 2.926 & --- &  \citet{2MASS}\\
HD 158774 & 4.403 & 3.451 & 3.138 & --- & \citet{2MASS}, \citet{Kawara1983}\\
HD 198611 & 3.755 & 2.862 & 2.470 & --- &  \citet{2MASS}, \citet{2MASS}, \citet{Neugebauer1969}\\
HD 202987 & 3.859 & 3.067 & 2.82 & 2.75 & \citet{2MASS}, \citet{Neugebauer1969}\\
3~Aqr & 0.934 & -0.020 & -0.220 & -0.338 & \citet{Carter1990}\\
HD 192909 & 1.190 & --- & 0.180 & 0.101 & \citet{Johnson1966}, \citet{Neugebauer1969}, \citet{Price1983}\\
$\xi$~Cyg & .995 & 0.130 & -0.070 & -0.150 & \citet{Johnson1966}, \citet{Noguchi1981}\\
HD 200817 & 4.174 & 3.721 & 3.708 & --- & \citet{2MASS}\\
HD 192464 & 5.180 & 4.176 & 3.879 & --- & \citet{2MASS}\\
$\alpha$~Cas & 0.371 & -0.191 & -0.270 & -0.399 & \citet{Voelcker1975}, \citet{Alonso1994}\\
$\mu$~Hya & 1.216 & .506 & 0.37 & 0.28 & \citet{2MASS}, \citet{Price1983}, \citet{Johnson1966}\\
HD 87262 & 2.974 & 2.052 & 1.880 & --- & \citet{2MASS}, \citet{Price1983}, \citet{Neugebauer1969}\\
HD 196321 & 2.128 & 1.361 & 1.21 & .98496 & \citet{2MASS}, \citet{Price1983}, \citet{Neugebauer1969}\\
HD 136422 & --- & --- & 0.8 & 0.535 &  \citet{Price1968}, \citet{Price1983}, \citet{Eggen1969} \\
$\delta$~Psc & 2.031 & 1.198 & .890 & 0.739 & \citet{2MASS}, \citet{Gosnell1979}\\
HD 198330 & 4.988 & 4.159 & 3.816 & --- & \citet{2MASS}\\
HD 188947 &  1.934 & 1.438 & 1.621 & 1.561 & \citet{Noguchi1981}, \citet{Elias1982}, \citet{Glass1975} \\
$\chi$~Cas & 3.019 & 2.481& 2.311 & --- & \citet{2MASS}, \citet{Neugebauer1969}\\
HD 163428 & --- & --- & 1.6 & 1.464& \citet{White1978}, \citet{Humphreys1974}\\ 
HD 196241 & 4.19 & 3.620 & 3.090 & ---& \citet{Morel1978}, \citet{2MASS}\\
\hline
\end{tabular}
\end{table*}

\begin{table*}
\tiny
\caption{Results from Circularly-Symmetric Gaussian Models}
\begin{tabular}{lllllll}
Target & Date(s) & Filter & Aperture
 & $V_0$ & FWHM & {$\chi$}$^2$/DOF\\
 & &&& ($\pm$0.05) & (mas) &\\
\hline
AFGL 230        & 1997 Dec  & k       & FFA          & 0.71 & 32$\pm$3 & 0.23 \\
                & 2002 Jul  & k       & FFA          & 0.54 & 34$\pm$3 & 0.34 \\
                &           & PAHcs   & FFA          & 0.74 & 33$\pm$2 & 0.05 \\
AFGL 2019       & 2000 Jun  & CH4     & annulus 36   & 0.96 & 10$^{+6}_{-10}$ & 0.31 \\
                &           & h       & annulus 36   & 0.90 & 9$\pm$4 & 0.65 \\
                &           & PAHcs   & annulus 36   & 0.95 & 21$\pm$3 & 0.27 \\
AFGL 2199       & 1998 Apr  & CH4     & annulus 36   & 0.92 & 14$\pm$6 & 0.23 \\
                &           & PAHcs   & annulus 36   & 1.00 & 22$\pm$3 & 0.45 \\
AFGL 2290*      & 1998 Jun  & CH4     & annulus 36   & 0.76 & 22$\pm$4 & 0.33 \\
                &           & PAHcs   & annulus 36   & 0.84 & 27$\pm$3 & 0.69 \\
                & 1999 Apr  & CH4     & annulus 36   & 0.72 & 34$\pm$3 & 0.39 \\
                &           & k       & annulus 36   & 0.75 & 32$\pm$3 & 0.51 \\
                &           & PAHcs   & annulus 36   & 0.83 & 36$\pm$2 & 0.36 \\
Cit 1*          & 2000 Jun  & CH4     & annulus 36   & 0.92 & 15$\pm$5 & 0.35 \\
                &           & h       & annulus 36   & 0.93 & 14$\pm$3 & 0.37 \\
                &           & PAHcs   & annulus 36   & 0.94 & 20$\pm$4 & 0.46 \\
Cit 3*          & 1997 Dec  & kcont   & annulus 36   & 0.89 & 20$\pm$5 & 0.44 \\
                &           & PAHcs   & annulus 36   & 0.89 & 37$\pm$2 & 0.21 \\
                & 1998 Sep  & CH4     & Golay 21     & 0.89 & 21$\pm$4 & 0.35 \\
                &           & PAHcs   & Golay 21     & 0.90 & 29$\pm$2 & 0.25 \\
v1300 Aql*      & 1998 Jun  & CH4     & annulus 36   & 0.83 & 14$\pm$6 & 0.43 \\
                &           & h       & annulus 36   & 0.81 & 14$\pm$3 & 0.41 \\
                &           & PAHcs   & annulus 36   & 0.84 & 23$\pm$3 & 0.50 \\
                & 1999 Jul  & kcont   & annulus 36   & 0.87 & 18$\pm$5 & 0.39 \\
                &           & PAHcs   & annulus 36   & 0.90 & 21$\pm$3 & 0.52 \\
\hline
AFGL 1922       & 2000 Jun  & k       & annulus 36   & 0.76 & 24$\pm$4 & 0.88 \\
                & 2001 Jun  & k       & annulus 36   & 0.83 & 29$\pm$4 & 0.76 \\
                &           & PAHcs   & annulus 36   & 0.95 & 58$\pm$2 & 0.43 \\
AFGL 1977*      & 1998 Jun  & CH4     & annulus 36   & 0.78 & 24$\pm$4 & 0.26 \\
                &           & h       & annulus 36   & 0.76 & 17$\pm$3 & 0.68 \\
                &           & PAHcs   & annulus 36   & 0.94 & 34$\pm$2 & 0.41 \\
                & 1999 Apr  & CH4     & annulus 36   & 0.96 & 29$\pm$4 & 0.25 \\
                &           & PAHcs   & annulus 36   & 0.89 & 52$\pm$2 & 0.26 \\
AFGL 2135       & 2001 Jun  & k       & annulus 36   & 0.66 & 17$\pm$5 & 0.49 \\
                & 2001 Jun  & PAHcs   & annulus 36   & 0.50 & 34$\pm$2 & 0.13 \\
AFGL 2232*      & 1998 Jun  & CH4     & annulus 36   & 0.83 & 18$\pm$5 & 1.32 \\
                &           & h       & annulus 36   & 0.81 & 14$\pm$3 & 0.50 \\
                &           & PAHcs   & annulus 36   & 0.90 & 33$\pm$2 & 0.56 \\
                &           & CH4     & Golay 21     & 0.91 & 20$\pm$5 & 0.17 \\
                &           & PAHcs   & Golay 21     & 0.90 & 34$\pm$2 & 0.14 \\
                & 1999 Apr  & CH4     & annulus 36   & 0.69 & 44$\pm$3 & 0.19 \\
                &           & PAHcs   & annulus 36   & 0.86 & 42$\pm$2 & 0.12 \\
AFGL 2513*      & 1998 Sep  & h       & annulus 36   & 1.00 & 1$^{+9}_{-1}$ & 1.15 \\
                &           & CH4     & annulus 36   & 0.94 & 10$^{+6}_{-10}$ & 0.18 \\
                &           & PAHcs   & annulus 36   & 1.00 & 16$\pm$4 & 0.70 \\
                & 1999 Jul  & CH4     & annulus 36   & 1.00 & 11$^{+6}_{-9}$ & 0.32 \\
                &           & PAHcs   & annulus 36   & 0.96 & 24$\pm$3 & 0.36 \\
AFGL 2686       & 1998 Sep  & CH4     & annulus 36   & 0.89 & 29$\pm$4 & 0.44 \\
                &           & h       & annulus 36   & 0.89 & 26$\pm$2 & 0.47 \\
                &           & PAHcs   & annulus 36   & 0.89 & 35$\pm$2 & 0.36 \\
                & 1999 Jul  & CH4     & annulus 36   & 0.92 & 26$\pm$4 & 0.68 \\
                &           & PAHcs   & annulus 36   & 0.91 & 33$\pm$2 & 0.44 \\
                &           & h       & annulus 36   & 0.77 & 28$\pm$2 & 0.87 \\
AFGL 4211       & 2000 Jun  & CH4     & annulus 36   & 0.78 & 31$\pm$3 & 0.48 \\
                &           & PAHcs   & annulus 36   & 0.82 & 70$\pm$3 & 0.10 \\
                & 2001 Jun  & k       & annulus 36   & 0.62 & 20$\pm$5 & 0.63 \\
IRAS~15148-4940 & 2001 Jun  & CH4     & annulus 36   & 0.77 & 13$\pm$7 & 0.41 \\
                &           & k       & annulus 36   & 0.82 & 13$\pm$7 & 0.59 \\
                &           & PAHcs   & annulus 36   & 0.86 & 25$\pm$3 & 0.39 \\
IY Hya          & 1999 Apr  & CH4     & annulus 36   & 0.88 & 14$\pm$6 & 0.32 \\
                &           & PAHcs   & annulus 36   & 0.94 & 33$\pm$2 & 0.28 \\
LP And*         & 1998 Sep  & CH4     & annulus 36   & 0.83 & 25$\pm$4 & 0.52 \\
                &           & h       & annulus 36   & 0.68 & 20$\pm$3 & 0.56 \\
                &           & PAHcs   & annulus 36   & 0.86 & 47$\pm$2 & 0.49 \\
                & 1999 Jul  & CH4     & Golay 21     & 0.89 & 24$\pm$4 & 0.99 \\
                &           & PAHcs   & Golay 21     & 0.79 & 48$\pm$2 & 0.50 \\
                & 1999 Jan  & CH4     & Golay 21     & 0.70 & 25$\pm$4 & 2.41 \\
                &           & PAHcs   & Golay 21     & 0.66 & 35$\pm$2 & 0.45 \\
RV Aqr*         & 1999 Jul  & CH4     & Golay 21     & 1.00 & 8$\pm$8 & 0.16 \\
                &           & PAHcs   & Golay 21     & 0.96 & 26$\pm$3 & 0.21 \\
                & 1998 Jun  & CH4     & Golay 21     & 0.98 & 12$\pm$8 & 0.13 \\
                &           & PAHcs   & Golay 21     & 1.00 & 27$\pm$3 & 0.36 \\
v1899 Cyg       & 1998 Jun  & CH4     & annulus 36   & 0.88 & 18$\pm$5 & 0.35 \\
                &           & h       & annulus 36   & 0.86 & 16$\pm$3 & 0.39 \\
                &           & PAHcs   & annulus 36   & 0.92 & 22$\pm$3 & 0.32 \\
                & 1999 Jul  & k       & annulus 36   & 0.93 & 15$\pm$5 & 0.62 \\
V Cyg           & 1998 Jun  & feii    & annulus 36   & 0.87 & 14$\pm$3 & 0.92 \\
                &           & kcont   & annulus 36   & 0.96 & 16$\pm$5 & 1.15 \\
                &           & PAHcs   & annulus 36   & 0.90 & 34$\pm$2 & 0.51 \\
                &           & CH4     & Golay 21     & 1.00 & 18$\pm$5 & 0.35 \\
                &           & PAHcs   & Golay 21     & 0.92 & 38$\pm$2 & 0.10 \\
                & 1999 Apr  & CH4     & Golay 21     & 0.93 & 17$\pm$5 & 0.15 \\
                &           & PAHcs   & Golay 21     & 0.86 & 38$\pm$2 & 0.12 \\
                & 2001 Jun  & CH4     & annulus 36   & 0.82 & 19$\pm$5 & 0.26 \\
                &           & PAHcs   & annulus 36   & 0.83 & 42$\pm$2 & 0.15 \\
\hline
\end{tabular}
\begin{flushleft}
* ~ target is asymmetric, see Table 7 for further details
\newline \textbf{Note:} Horizontal line separates oxygen-rich (top) from carbon-rich (bottom).
\end{flushleft}
\end{table*}

\begin{table*}
\tiny
\caption{Results from Central Point plus Circularly-Symmetric Gaussian Models}
\begin{tabular}{llllllll}
Target & Date(s)  & Filter  & Aperture
 & f$_{\rm Point}$  & f$_{\rm Gauss}$  &  FWHM & $\chi^2$/DOF\\
           &  &  & & ($\pm$.05) & ($\pm$0.05) & (mas) & \\
\hline
AFGL 230        & 1997 Dec  & k       & FFA          & 0.24 & 0.52 & 47$\pm$7 & 0.21 \\
                & 2002 Jul  & k       & FFA          & 0.30 & 0.42 & 98$\pm$7 & 0.15 \\
                &           & PAHcs   & FFA          & 0.50 & 0.32 & 86$\pm$8 & 0.42 \\
AFGL 2019       & 2000 Jun  & CH4     & annulus 36   & 0.86 & 0.14 & 51$^{+30}_{-45}$ & 0.28 \\
                &           & h       & annulus 36   & 0.00 & 0.83 & 1$^{+9}_{-1}$ & 0.83 \\
                &           & PAHcs   & annulus 36   & 0.45 & 0.51 & 31$\pm$5 & 0.27 \\
AFGL 2199       & 1998 Apr  & CH4     & annulus 36   & 0.38 & 0.54 & 19$\pm$9 & 0.22 \\
                &           & PAHcs   & annulus 36   & 0.36 & 0.68 & 30$\pm$4 & 0.43 \\
AFGL 2290*       & 1998 Jun  & CH4     & annulus 36   & 0.44 & 0.38 & 46$\pm$12 & 0.26 \\
                &           & PAHcs   & annulus 36   & 0.55 & 0.38 & 68$\pm$7 & 0.61 \\
                & 1999 Apr  & CH4     & annulus 36   & 0.31 & 0.51 & 66$\pm$9 & 0.21 \\
                &           & k       & annulus 36   & 0.34 & 0.56 & 66$\pm$9 & 0.31 \\
                &           & PAHcs   & annulus 36   & 0.25 & 0.60 & 49$\pm$3 & 0.35 \\
CIT 1*           & 2000 Jun  & CH4     & annulus 36   & 0.00 & 0.92 & 14$\pm$6 & 0.35 \\
                &           & h       & annulus 36   & 0.47 & 0.49 & 23$\pm$6 & 0.36 \\
                &           & PAHcs   & annulus 36   & 0.00 & 0.94 & 20$\pm$4 & 0.46 \\
CIT 3  *         & 1997 Dec  & kcont   & annulus 36   & 0.58 & 0.50 & 53$\pm$13 & 0.23 \\
                &           & PAHcs   & annulus 36   & 0.36 & 0.62 & 60$\pm$4 & 0.02 \\
                & 1998 Sep  & CH4     & Golay 21     & 0.50 & 0.44 & 40$\pm$10 & 0.03 \\
                &           & PAHcs   & Golay 21     & 0.46 & 0.47 & 50$\pm$5 & 0.20 \\
v1300 Aql*       & 1998 Jun  & CH4     & annulus 36   & 0.64 & 0.23 & 43$\pm$20 & 0.40 \\
                &           & h       & annulus 36   & 0.45 & 0.38 & 25$\pm$7 & 0.39 \\
                &           & PAHcs   & annulus 36   & 0.64 & 0.34 & 80$\pm$10 & 0.30 \\
                & 1999 Jul  & kcont   & annulus 36   & 0.00 & 0.87 & 18$\pm$5 & 0.39 \\
                &           & PAHcs   & annulus 36   & 0.17 & 0.73 & 24$\pm$4 & 0.52 \\
\hline
AFGL 1922       & 2000 Jun  & k       & annulus 36   & 0.42 & 0.39 & 47$\pm$11 & 0.83 \\
                & 2001 Jun  & k       & annulus 36   & 0.43 & 0.49 & 57$\pm$12 & 0.62 \\
                &           & PAHcs   & annulus 36   & 0.51 & 0.47 & 105$\pm$6 & 0.41 \\
AFGL 1977*       & 1998 Jun  & CH4     & annulus 36   & 0.36 & 0.45 & 41$\pm$9 & 0.22 \\
                &           & h       & annulus 36   & 0.50 & 0.33 & 43$^{+8}_{-13}$ & 0.59 \\
                &           & PAHcs   & annulus 36   & 0.42 & 0.58 & 56$\pm$4 & 0.36 \\
                & 1999 Apr  & CH4     & annulus 36   & 0.34 & 0.69 & 43$\pm$7 & 0.17 \\
                &           & PAHcs   & annulus 36   & 0.25 & 0.74 & 45$\pm$3 & 0.17 \\
AFGL 2135       & 2001 Jun  & k       & annulus 36   & 0.00 & 0.66 & 17$\pm$5 & 0.49 \\
                & 2001 Jun  & PAHcs   & annulus 36   & 0.28 & 0.27 & 78$\pm$6 & 0.10 \\
AFGL 2232*       & 1998 Jun  & CH4     & annulus 36   & 0.56 & 0.33 & 44$\pm$14 & 0.47 \\
                &           & h       & annulus 36   & 0.00 & 0.81 & 14$\pm$3 & 1.32 \\
                &           & PAHcs   & annulus 36   & 0.50 & 0.49 & 66$\pm$6 & 0.49 \\
                &           & CH4     & Golay 21     & 0.52 & 0.45 & 39$\pm$10 & 0.10 \\
                &           & PAHcs   & Golay 21     & 0.35 & 0.60 & 51$\pm$4 & 0.09 \\
                & 1999 Apr  & CH4     & annulus 36   & 0.23 & 0.56 & 72$\pm$7 & 0.08 \\
                &           & PAHcs   & annulus 36   & 0.26 & 0.64 & 58$\pm$3 & 0.07 \\
AFGL 2513*       & 1998 Sep  & h       & annulus 36   & 0.03 & 1.00 & 1$^{+9}_{-1}$ & 1.09 \\
                &           & CH4     & annulus 36   & 0.81 & 0.14 & 39$\pm$39 & 0.17 \\
                &           & PAHcs   & annulus 36   & 0.03 & 1.00 & 19$\pm$4 & 0.68 \\
                & 1999 Jul  & CH4     & annulus 36   & 0.00 & 0.92 & 1$^{+13}_{-1}$ & 0.65 \\
                &           & PAHcs   & annulus 36   & 0.67 & 0.37 & 65$\pm$9 & 0.24 \\
AFGL 2686       & 1998 Sep  & CH4     & annulus 36   & 0.30 & 0.63 & 42$\pm$7 & 0.39 \\
                &           & h       & annulus 36   & 0.24 & 0.69 & 35$\pm$4 & 0.43 \\
                &           & PAHcs   & annulus 36   & 0.39 & 0.56 & 58$\pm$4 & 0.29 \\
                & 1999 Jul  & CH4     & annulus 36   & 0.21 & 0.73 & 32$\pm$5 & 0.67 \\
                &           & PAHcs   & annulus 36   & 0.29 & 0.64 & 45$\pm$3  & 0.42 \\
                &           & h       & annulus 36   & 0.63 & 0.30 & 138$^{+15}_{-9}$ & 0.77 \\
AFGL 4211       & 2000 Jun  & CH4     & annulus 36   & 0.36 & 0.54 & 59$\pm$10 & 0.27 \\
                &           & PAHcs   & annulus 36   & 0.16 & 0.70 & 89$\pm$4 & 0.06 \\
                & 2001 Jun  & k       & annulus 36   & 0.47 & 0.28 & 86$^{+9}_{-19}$ & 0.45 \\
IRAS~15148-4940 & 2001 Jun  & CH4     & annulus 36   & 0.62 & 0.18 & 44$\pm$23 & 0.39 \\
                &           & k       & annulus 36   & 0.52 & 0.31 & 25$\pm$13 & 0.59 \\
                &           & PAHcs   & annulus 36   & 0.52 & 0.38 & 49$\pm$7 & 0.37 \\
IY Hya          & 1999 Apr  & CH4     & annulus 36   & 0.00 & 0.88 & 14$\pm$6 & 0.32 \\
                &           & PAHcs   & annulus 36   & 0.00 & 0.94 & 32$\pm$2 & 0.28 \\
LP And*          & 1998 Sep  & CH4     & annulus 36   & 0.42 & 0.49 & 49$\pm$10 & 0.41 \\
                &           & h       & annulus 36   & 0.37 & 0.34 & 39$\pm$9 & 0.52 \\
                &           & PAHcs   & annulus 36   & 0.30 & 0.65 & 74$\pm$4 & 0.35 \\
                & 1999 Jul  & CH4     & Golay 21     & 0.47 & 0.49 & 47$\pm$11 & 0.91 \\
                &           & PAHcs   & Golay 21     & 0.35 & 0.51 & 85$\pm$5 & 0.42 \\
                & 1999 Jan  & CH4     & Golay 21     & 0.47 & 0.49 & 47$\pm$11 & 0.91 \\
                &           & PAHcs   & Golay 21     & 0.35 & 0.51 & 85$\pm$5 & 0.42 \\
RV Aqr*          & 1999 Jul  & CH4     & Golay 21     & 0.00 & 0.96 & 1$^{+13}_{-1}$ & 0.30 \\
                &           & PAHcs   & Golay 21     & 0.33 & 0.64 & 34$\pm$4 & 0.21 \\
                & 1998 Jun  & CH4     & Golay 21     & 0.74 & 0.25 & 30$\pm$20 & 0.12 \\
                &           & PAHcs   & Golay 21     & 0.43 & 0.62 & 42$\pm$4 & 0.33 \\
v1899 Cyg       & 1998 Jun  & CH4     & annulus 36   & 0.32 & 0.57 & 24$\pm$7 & 0.35 \\
                &           & h       & annulus 36   & 0.72 & 0.23 & 75$\pm$15 & 0.30 \\
                &           & PAHcs   & annulus 36   & 0.65 & 0.23 & 57$\pm$13 & 0.31 \\
                & 1999 Jul  & k       & annulus 36   & 0.76 & 0.23 & 52$\pm$24 & 0.59 \\
V Cyg           & 1998 Jun  & feii    & annulus 36   & 0.31 & 0.56 & 18$\pm$5 & 0.91 \\
                &           & kcont   & annulus 36   & 0.51 & 0.47 & 26$\pm$9 & 1.14 \\
                &           & PAHcs   & annulus 36   & 0.00 & 0.90 & 34$\pm$2 & 0.51 \\
                &           & CH4     & Golay 21     & 0.58 & 0.47 & 35$\pm$10 & 0.30 \\
                &           & PAHcs   & Golay 21     & 0.28 & 0.67 & 52$\pm$3 & 0.06 \\
                & 1999 Apr  & CH4     & Golay 21     & 0.52 & 0.43 & 30$\pm$9 & 0.14 \\
                &           & PAHcs   & Golay 21     & 0.31 & 0.60 & 57$\pm$4 & 0.07 \\
                & 2001 Jun  & CH4     & annulus 36   & 0.50 & 0.36 & 40$\pm$11 & 0.21 \\
                &           & PAHcs   & annulus 36   & 0.28 & 0.61 & 63$\pm$4 & 0.09 \\
\hline
\end{tabular}
\begin{flushleft}
* ~ target is asymmetric, see Table 7 for further details\\
\textbf{Note:} Horizontal line separates oxygen-rich (top) from carbon-rich (bottom).
\end{flushleft}
\end{table*}

\begin{table*}
\caption{Results from 2-dimensional Gaussian Models}
\begin{tabular}{lllll}
Target & Date(s)  & Filter &  $\frac{{\rm FWHM}_{\rm major}} {{\rm FWHM}_{\rm minor}}$ & $<$PA$>$\\
\hline
AFGL 2290 & 1998 Jun      & CH4      & 1.23 & 58$\pm$20\\ 
                   &                       & PAHcs   & 1.24 & \\ 
                   & 1999 Apr     & CH4        & 1.24 & \\ 
                   &                       & k            & 1.24 & \\
                   &                       & PAHcs   & 1.06 & \\  
CIT 1        & 2000 Jun & CH4  & 1.13 & 133$\pm$3\\
                &                    & h        & 1.14 &\\
                &			& PAHcs & 1.11 & \\
CIT 3        & 1997 Dec & Kcont  & 1.19  & 151$\pm$9\\
	       &                    & PAHcs  & 1.06 & \\
	       & 1998 Sep &  CH4      & 1.28  & \\ 
	       &                    & PAHcs  & 1.04 & \\
v1300 Aql & 1998 Jun    & CH4      & 1.34 & 108$\pm$13\\
                   &                      & h          & 1.14 & \\
                   &                      & PAHcs & 1.19 & \\
                   & 1999 Jul      & kcont  & 1.34 & \\
                   &                       & PAHcs & 1.31& \\
\hline
AFGL 1977 & 1998 Jun      & CH4      & 1.11 & 71$\pm$18\\
                  &                        & h          & 1.31  & \\   
                  &                        & PAHcs  & 1.06 & \\
                  & 1999 Apr      & CH4       & 1.10  & \\
                  &                        & PAHcs  & 1.21 & \\
AFGL 2232  & 1998 Jun     & CH4       & 1.37 & 94$\pm$10\\
                    &                      & h            & 1.83 & \\
                    &                       &PAHcs   & 1.19 & \\
                    &                      & CH4        & 1.10 & \\
                    &                       & PAHcs   & 1.08 & \\
                    & 1999 Apr     & CH4        & 1.22 & \\
                    &                       & PAHcs   & 1.05  & \\
AFGL 2513 & 1998 Sep  & h          & unresolved & \\ 
                  &                     & CH4      &1.5 & 61$\pm$8\\
                   &                    & PAHcs & 1.39 & \\
                  & 1999 Jul    & CH4      & 1.38 & \\
                  &                     & PAHcs & 1.2 &  \\
LP And     & 1998 Sep   & CH4      & 1.46 & 108$\pm$6\\
                  &                     & h           & 2.03 & \\
                  &                     & PAHcs  & 1.39 & \\
                  & 1999 Jul     & CH4      & 1.64 & \\
                  &                      & PAHcs & 1.36  & \\
                  & 1999 Jan    & CH4      & 1.82 & \\
                  &                      & PAHcs  & 1.20 & \\
RV Aqr           & 1999 Jul      & CH4       & 1.49 & 122$\pm$24\\ 
                     &                       & PAHcs    & 1.23 & \\ 
                     & 1998 Jun     & CH4        & 1.35 & \\
                     &                       & PAHcs    & 1.22 & \\
\hline
\end{tabular}
\begin{flushleft}
* Sources missing from this list were found to have circularly-symmetric dust shells (within errors).
\textbf{Note:} PA is the mean position angle of the major axis (degrees East of North) for all filters and epochs.
\end{flushleft}
\end{table*}

\begin{table*}
\caption{Results from DUSTY Radiative Transfer Model}
\begin{tabular}{lllllll}
 Target & Date(s) & T$_{\rm dust}$  & $\tau_{2.2\mu\,m}$ & R$_\ast$ & L$_\ast^{(1\,kpc)}$ & $\chi^2$/DOF \\
              &               & (K)  &  & (mas) &$(10^3$~L$_\odot$) &\\
\hline
AFGL 230   & 1997 Dec & 800$^{+60}_{-90}$ & 4.9$^{+0.9}_{-0.7}$ & 1.5$^{+0.5}_{-0.3}$ & 4.5$^{+3.1}_{-1.6}$ & 0.26\\
                  & 2002 Jul   & 540$^{+400}_{-110}$ & 7.4$^{+1.6}_{-1.2}$ & 4.1$^{+4.5}_{-2.1}$ & 31$^{+108}_{-24}$ & 3.86 \\
AFGL 2019 & 2000 Jun  &1190$^{+310}_{-250}$ & 0.92$^{+0.23}_{-0.12}$ & 3.5$^{+0.7}_{-0.3}$ & 24$^{+10}_{-4}$ & 0.54\\
AFGL 2199 & 1998 Apr & 1130$^{+370}_{-310}$ & 1.6$^{+1.2}_{-0.7}$ & 3.3$^{+2.0}_{-0.6}$ & 21$^{+32}_{-7}$ & 0.06\\
AFGL 2290 & 1998 Jun  & 850$^{+140}_{-80}$ & 3.5$^{+0.5}_{-0.5}$ & 3.7$^{+0.8}_{-0.7}$ & 26$^{+12}_{-9}$ & 0.33\\ 
                   & 1999 Apr  & 800$^{+140}_{-140}$ & 4.6$^{+0.7}_{-0.5}$ & 3.9$^{+1.7}_{-0.8}$ & 29$^{+31}_{-11}$ & 2.63\\ 
CIT 1        & 2000 Jun & 1190$^{+310}_{-230}$ & 1.2$^{+0.5}_{-0.2}$ & 3.5$^{+0.8}_{-0.4}$ & 24$^{+12}_{-5}$ & 0.60\\
CIT 3        & 1997 Dec & 1110$^{+230}_{-140}$ & 1.4$^{+0.7}_{-0.5}$ & 7.8$^{+1.5}_{-0.6}$ & 116$^{+49}_{-16}$ & 0.34\\
	       & 1998 Sep  & 1020$^{+200}_{-110}$ & 1.9$^{+0.7}_{-0.5}$ & 5.0$^{+0.8}_{-0.6}$ & 48$^{+17}_{-11}$ & 0.29\\ 
v1300 Aql & 1998 Jun  & 1080$^{+340}_{-170}$ & 0.92$^{+0.46}_{-0.12}$ & 5.8$^{+1.0}_{-0.6}$ & 64$^{+24}_{-12}$ & 0.50\\
                   & 1999 Jul   & 1160$^{+340}_{-250}$* & 1.60$^{+0.9}_{-0.7}$* & 5.1$^{+2.0}_{-0.8}$ & 49$^{+45}_{-14}$ & 0.11\\
\hline
AFGL 1922 & 2000 Jun  & 850$^{+200}_{-60}$ & 5.3$^{+0.7}_{-0.7}$ & 4.5$^{+1.1}_{-0.8}$ & 37$^{+21}_{-13}$ & 1.44\\
                   & 2001 Jun  & 850$^{+170}_{-60}$ & 3.9$^{+0.5}_{-0.5}$ & 5.2$^{+1.2}_{-1.5}$ & 51$^{+27}_{-25}$ & 0.39\\
AFGL 1977 & 1998 Jun   & 910$^{+80}_{-90}$ & 2.8$^{+0.1}_{-0.2}$ & 4.0$^{+0.2}_{-0.6}$ & 31$^{+4}_{-8}$ & 1.69\\
                  & 1999 Apr    & 990$^{+90}_{-60}$ & 2.5$^{+0.5}_{-0.2}$ & 6.0$^{+0.8}_{-0.4}$ & 68$^{+19}_{-9}$ & 0.65\\
AFGL 2135   & 2001 Jun & 740$^{+370}_{-200}$ & 3.2$^{+4.2}_{-1.4}$ & 9.4$^{+65.7}_{-4.6}$ & 167$^{+10500}_{-123}$ & 5.04\\
AFGL 2232  & 1998 Jun  & 1110$^{+140}_{-110}$ & 1.6$^{+0.2}_{-0.2}$ & 5.1$^{+0.5}_{-0.5}$ & 48$^{+11}_{-10}$ & 0.32\\
                    & 1999 Apr  & 1300$^{+200}_{-230}$ & 2.8$^{+1.4}_{-1.2}$ & 9.4$^{+5.0}_{-1.2}$ & 165$^{+226}_{-41}$ & 3.29\\
AFGL 2513 & 1998 Sep   & 1500$^{+0}_{-450}$ & 1.9$^{+0.2}_{-0.2}$ & 1.7$^{+0.9}_{-0.2}$ & 5.3$^{+7.3}_{-0.9}$ & 0.57\\ 
                  & 1999 Jul     & 1110$^{+400}_{-200}$ & 1.2$^{+0.9}_{-0.5}$ & 3.1$^{+1.0}_{-0.4}$ & 18$^{+14}_{-4}$ & 0.30\\
AFGL 2686 & 1998 Sep   & 1110$^{+140}_{-140}$ & 2.8$^{+0.5}_{-0.5}$ & 5.3$^{+1.1}_{-0.8}$ & 53$^{+24}_{-15}$ & 1.85\\
                  & 1999 Jul     & 820$^{+60}_{-60}$ & 3.2$^{+0.2}_{-0.2}$ & 3.1$^{+0.4}_{-0.3}$ & 19$^{+4}_{-4}$ & 1.02\\
AFGL 4211  & 2000 Jun  & 880$^{+60}_{-30}$ & 3.7$^{+0.7}_{-0.5}$ & 6.7$^{+0.3}_{-0.3}$ & 85$^{+9}_{-7}$ & 0.73\\
                   & 2001 Jun  & 850$^{+170}_{-80}$ & 4.2$^{+0.9}_{-0.7}$ & 6.1$^{+3.7}_{-2.0}$ & 71$^{+113}_{-39}$ & 3.20\\
IRAS~15148-4940 & 2001 Jun  & 940$^{+340}_{-170}$* & 0.23$^{+0.46}_{-0.12}$ & 4.6$^{+0.2}_{-0.9}$ & 40$^{+4}_{-14}$ & 2.47\\    
IY Hya          & 1999 Apr  & 960$^{+140}_{-110}$ & 0.46$^{+0.46}_{-0.12}$ & 4.2$^{+0.1}_{-0.2}$ & 33$^{+2}_{-3}$ & 0.25\\
LP And     & 1998 Sep   & 880$^{+70}_{-60}$ & 3.0$^{+0.5}_{-0.2}$ & 4.8$^{+0.9}_{-0.5}$ & 43$^{+19}_{-8}$ & 1.49\\
                  & 1999 Jul     & 820$^{+60}_{-60}$ & 3.0$^{+0.2}_{-0.5}$ & 4.8$^{+0.6}_{-0.7}$ & 44$^{+12}_{-12}$ & 1.00\\
                  & 1999 Jan    & 880$^{+90}_{-30}$ & 3.2$^{+0.5}_{-0.2}$ & 6.7$^{+1.0}_{-0.7}$ & 85$^{+27}_{-17}$ & 1.67\\
RV Aqr           & 1999 Jul   & 1500$^{+0}_{-280}$ & 0.46$^{+0.23}_{-0.23}$ & 5.4$^{+0.7}_{-0.4}$ & 55$^{+15}_{-7}$ & 1.09\\
                     & 1998 Jun  & 1190$^{+310}_{-150}$ & 0.23$^{+0.46}_{-0.12}$ & 5.1$^{+0.1}_{-0.7}$ & 48$^{+2}_{-13}$ & 0.23\\
v1899 Cyg & 1998 Jun   & 740$^{+60}_{-60}$ & 2.3$^{+0.5}_{-0.2}$ & 2.0$^{+0.5}_{-0.3}$ & 7.5$^{+3.8}_{-1.8}$ & 0.53\\
                     & 1999 Jul    & 600$^{+340}_{-200}$ & 2.5$^{+2.5}_{-1.2}$ & 1.8$^{+4.4}_{-1.0}$ & 6.1$^{+66}_{-4.8}$ & 0.12\\
V Cyg          & 1998 Jun    & 1270$^{+230}_{-200}$* & 0.69$^{+0.46}_{-0.23}$* & 6.6$^{+0.7}_{-0.1}$ & 83$^{+19}_{-3}$ & 3.02\\
                   & 1999 Apr     & 1160$^{+200}_{-110}$ & 0.23$^{+0.23}_{-0.12}$ & 9.6$^{+0.3}_{-1.2}$ & 174$^{+11}_{-40}$ & 0.21\\
                   & 2001 Jun     & 1270$^{+140}_{-140}$ & 0.46$^{+0.23}_{-0.23}$ & 10.6$^{+1.6}_{-0.6}$ & 212$^{+70}_{-24}$ & 0.24\\
\hline
\end{tabular}
\begin{flushleft}
* ~ This star has two regions which meet our 1-$\sigma$ criteria for a best fit.  The particular values 
shown were chosen for consistency, see the appropriate figure for more details.
\newline \textbf{Note:} Horizontal line separates oxygen-rich (top) from carbon-rich (bottom).
\end{flushleft}
\end{table*}

\clearpage

\begin{figure}
\centering
\includegraphics[angle=90,width=3in]{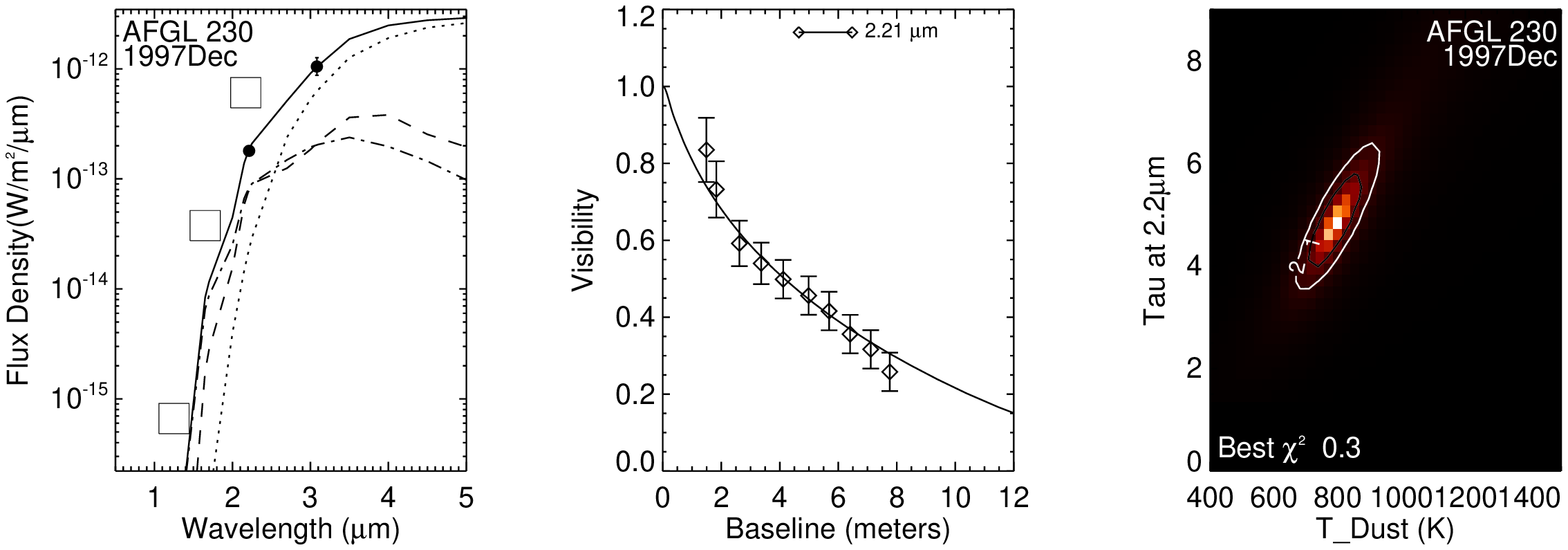} 
\includegraphics[angle=90,width=3in]{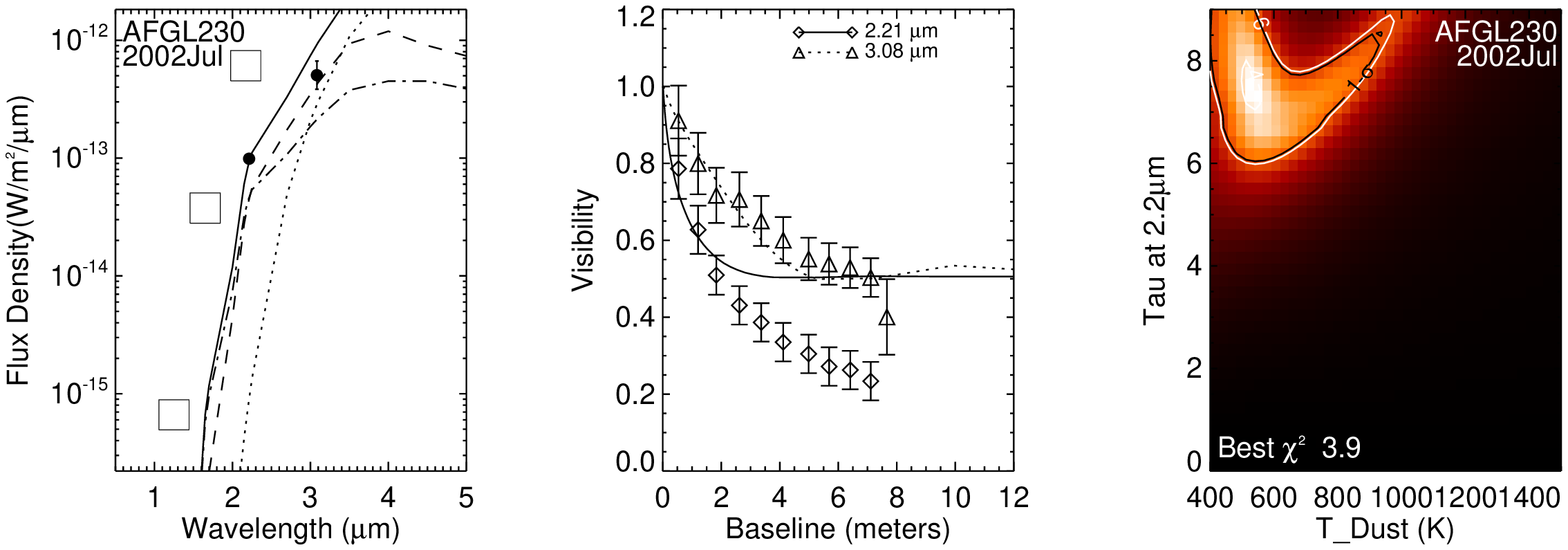}
\caption{Best fit plots for AFGL230.  The first row are figures for the epoch Dec97 and 
the second row is for Jul02.  The first panel in each row shows a fit to the SED with our new photometry included with errors (2MASS points are plotted as squares in each frame for reference).  The dashed line represents 
the contribution from the star, the dotted line represents dust contribution, the dash-dotted line represents
 the contribution from scattered light, and the solid line is the total flux.  The second panel shows our DUSTY fits to the visibility data for each wavelength of observations.   The third panel shows the $\chi^2/DOF$ surface, with bright areas showing the best-fitting region. The black contour denotes the 1-$\{\sigma\}$ error .}
\end{figure}

\begin{figure}
\centering
\includegraphics[angle=90,width=3in]{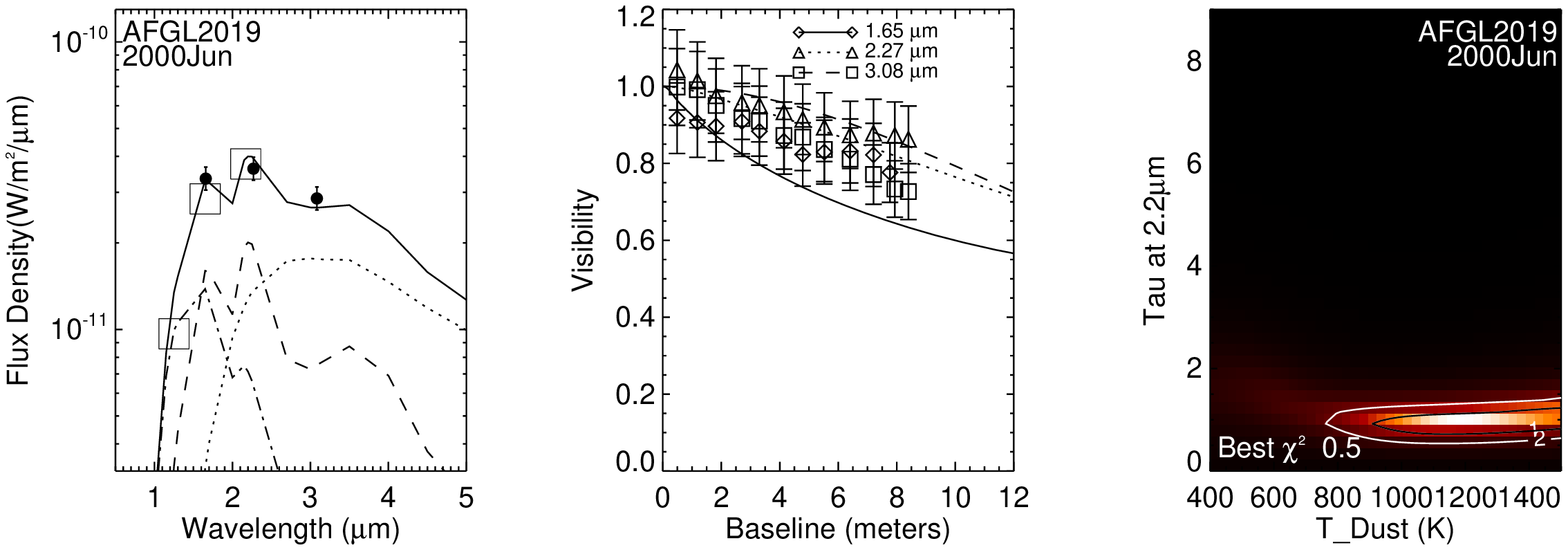}
\caption{ Best fit plots for AFGL2019.  See Fig.1 caption.}
\end{figure}

\begin{figure}
\centering
\includegraphics[angle=90,width=3in]{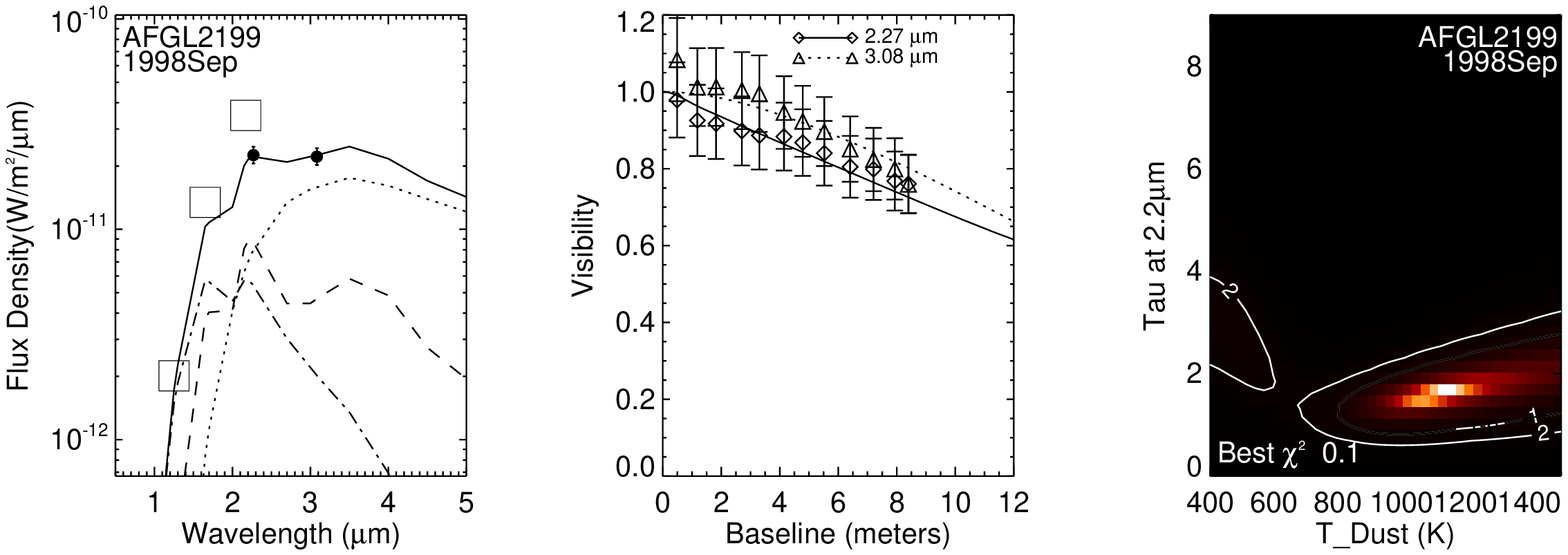}
\caption{ Best fit plots for AFGL2199.  See Fig.1 caption.}
\end{figure}

\begin{figure}
\centering
\includegraphics[angle=90,width=3in]{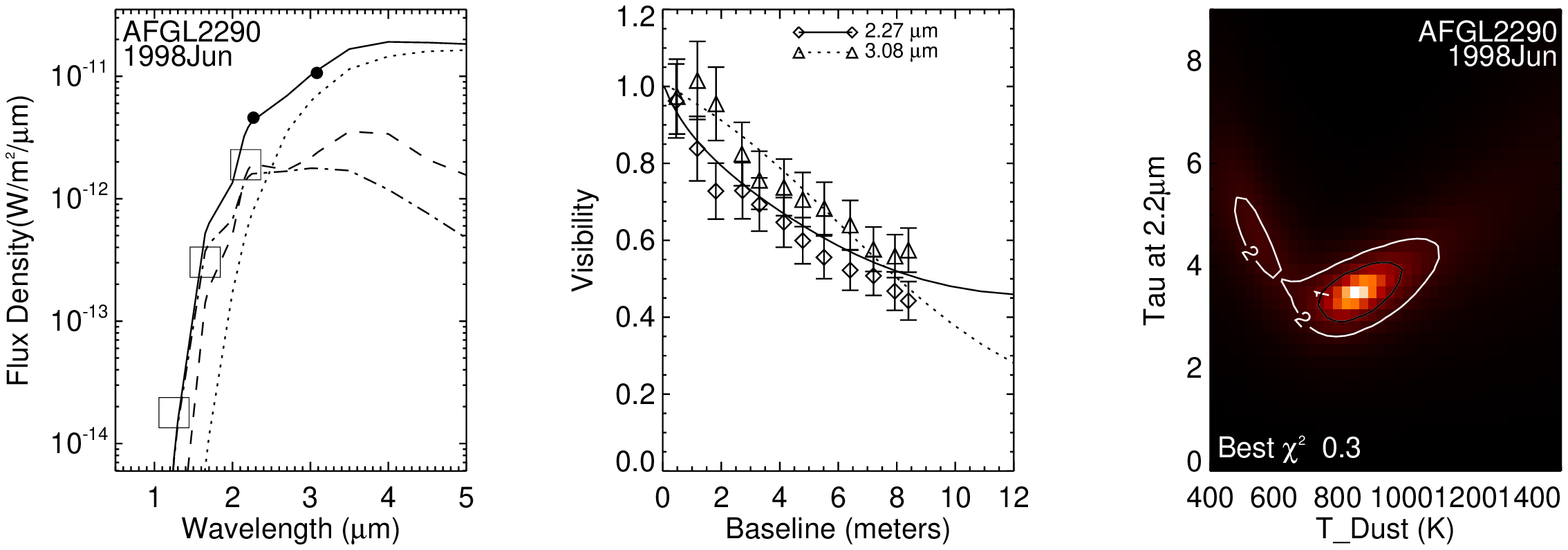}
\includegraphics[angle=90,width=3in]{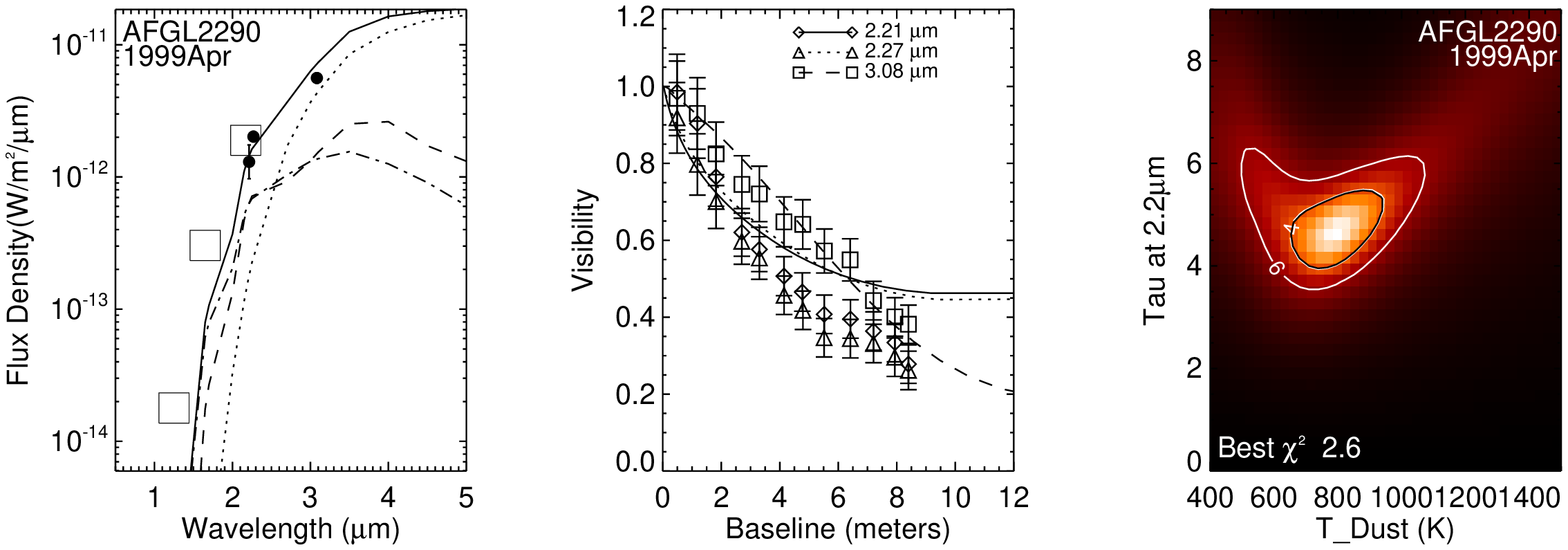}
\caption{ Best fit plots for AFGL2290.  See Fig.1 caption.}
\end{figure}

\begin{figure}
\centering
\includegraphics[angle=90,width=3in]{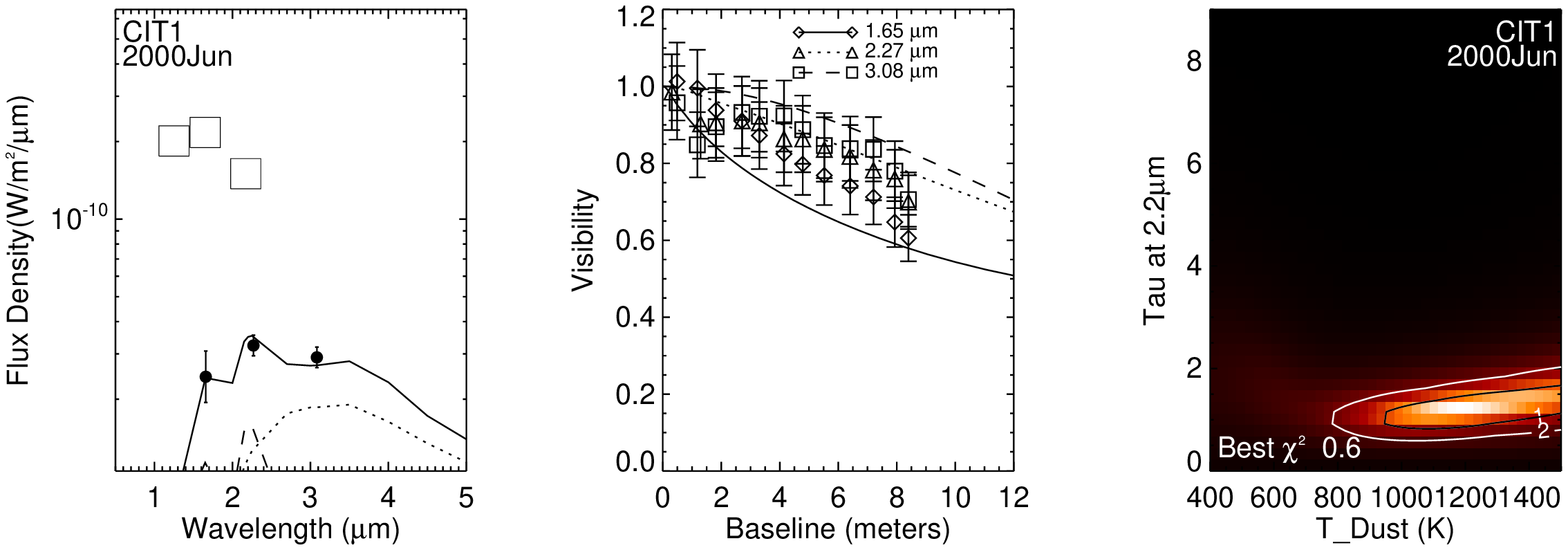}
\caption{ Best fit plots for CIT~1.  See Fig.1 caption.}
\end{figure}

\begin{figure}
\centering
\includegraphics[angle=90,width=3in]{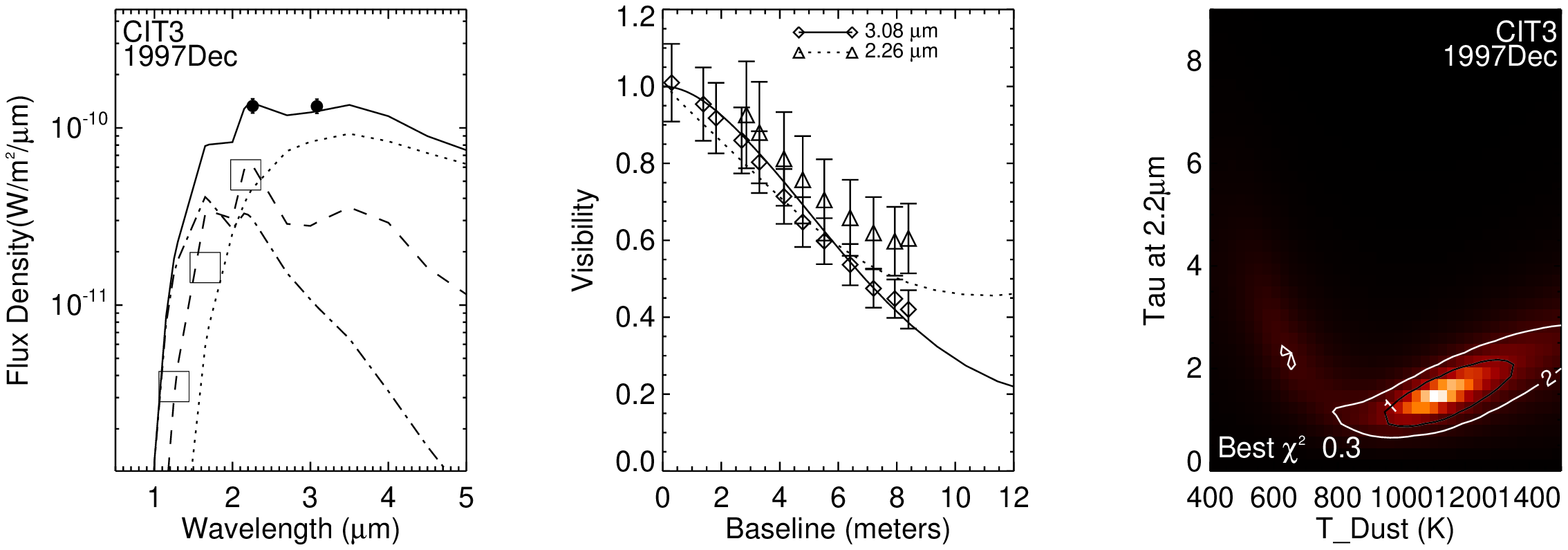}
\includegraphics[angle=90,width=3in]{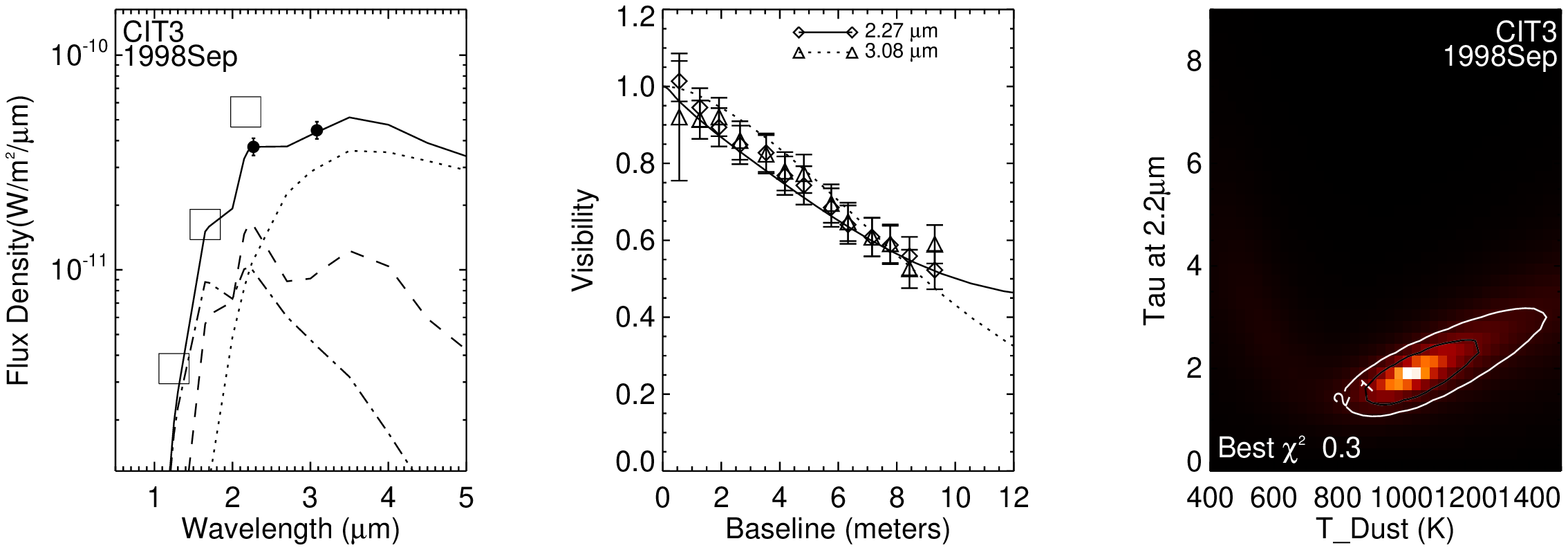}
\caption{ Best fit plots for CIT~3.  See Fig.1 caption.}
\end{figure}

\begin{figure}
\centering
\includegraphics[angle=90,width=3in]{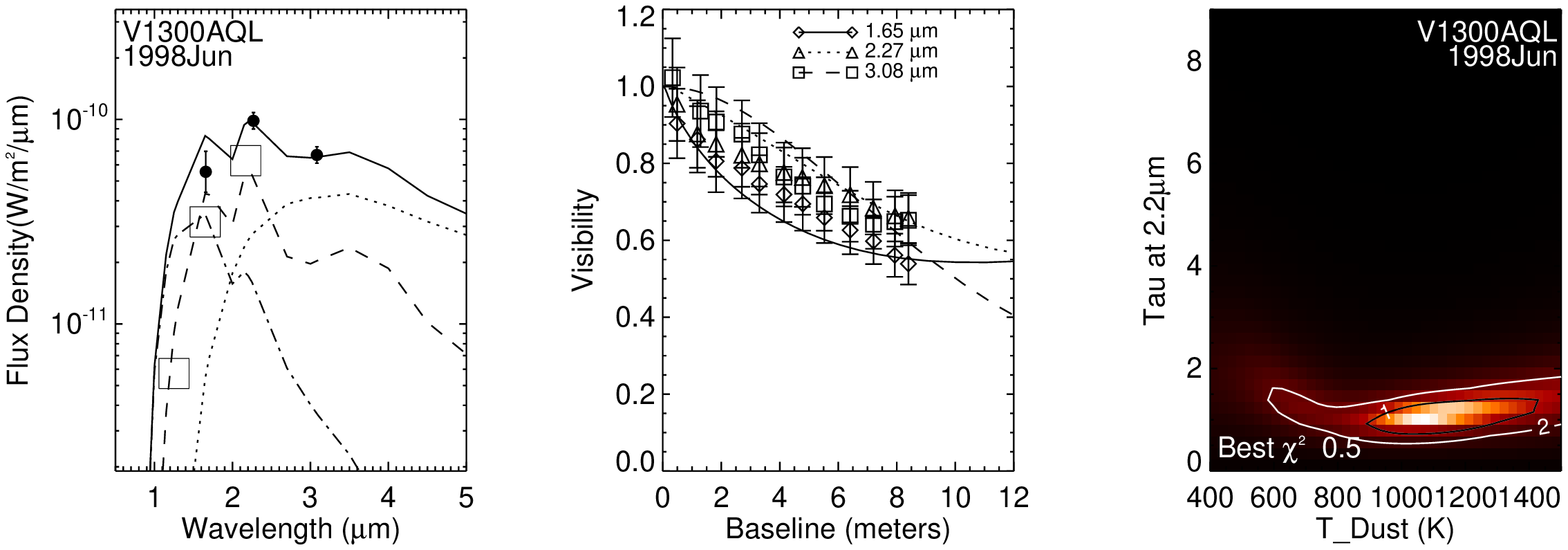}
\includegraphics[angle=90,width=3in]{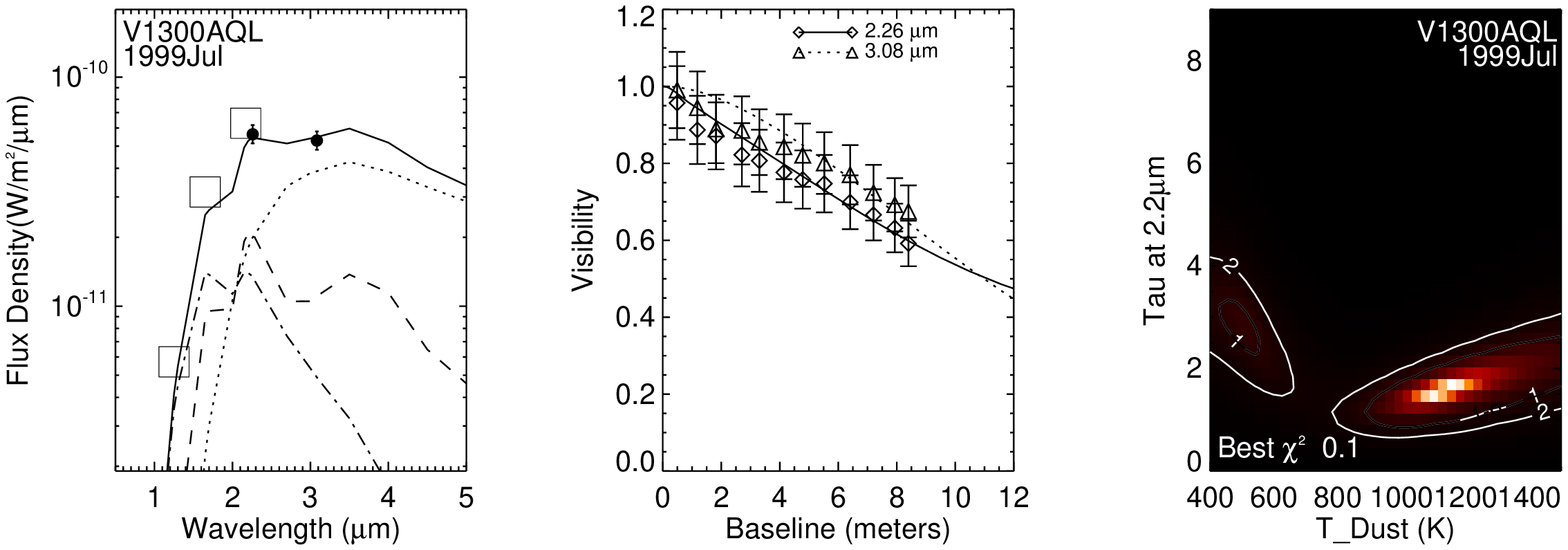}
\caption{ Best fit plots for v1300 Aql.  See Fig.1 caption.  For the 1999 Jul. epoch
we chose the lower-right region as the best fit region because it is consistent with the best fit region
for the 1998 Jun. epoch.}
\end{figure}

\begin{figure}
\centering
\includegraphics[angle=90,width=3in]{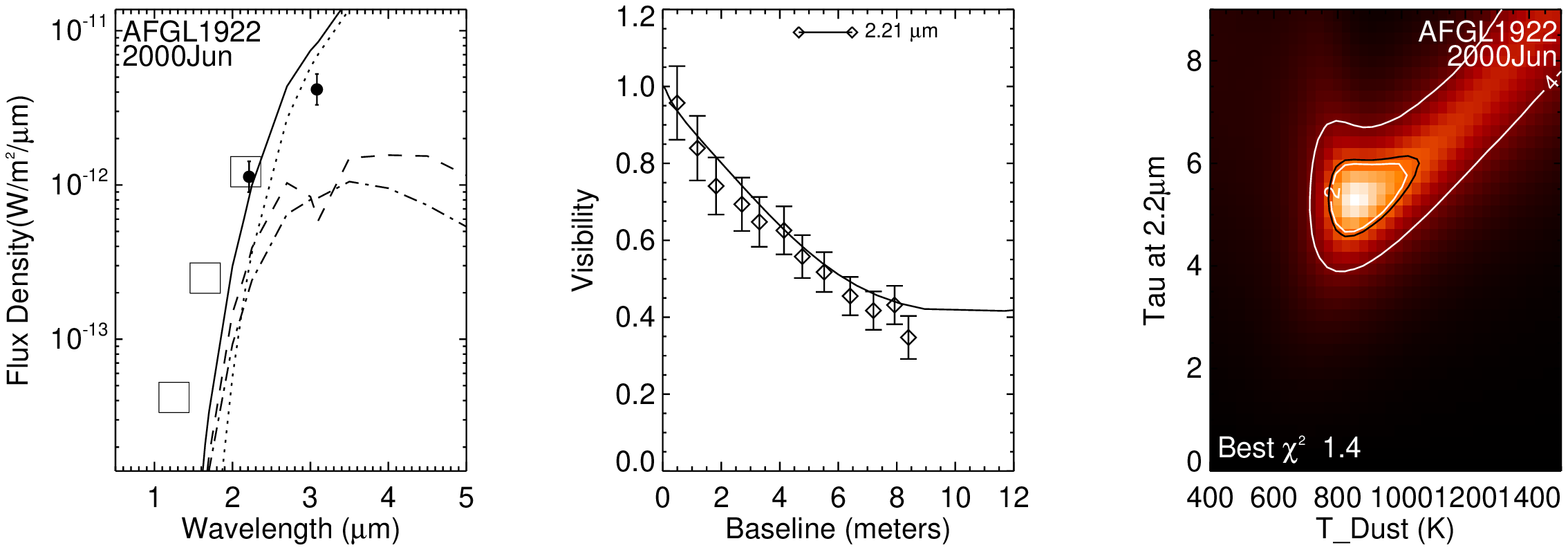}
\includegraphics[angle=90,width=3in]{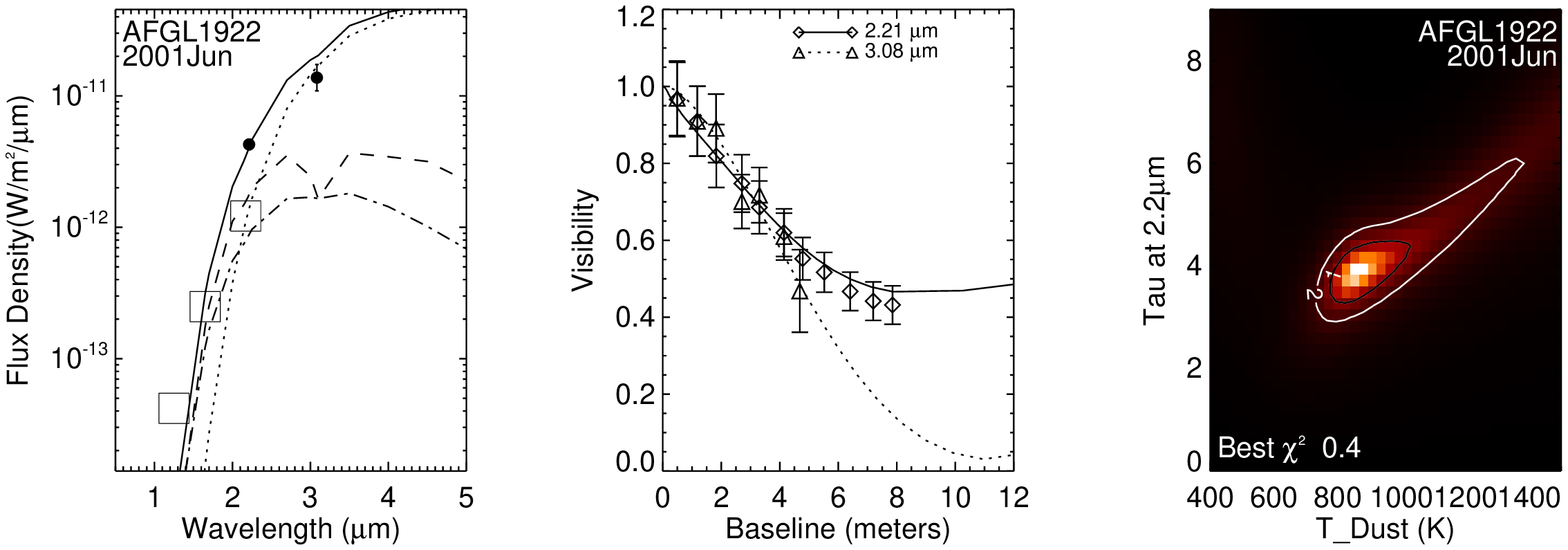}
\caption{ Best fit plots for AFGL1922.  See Fig.1 caption.}
\end{figure}

\begin{figure}
\centering
\includegraphics[angle=90,width=3in]{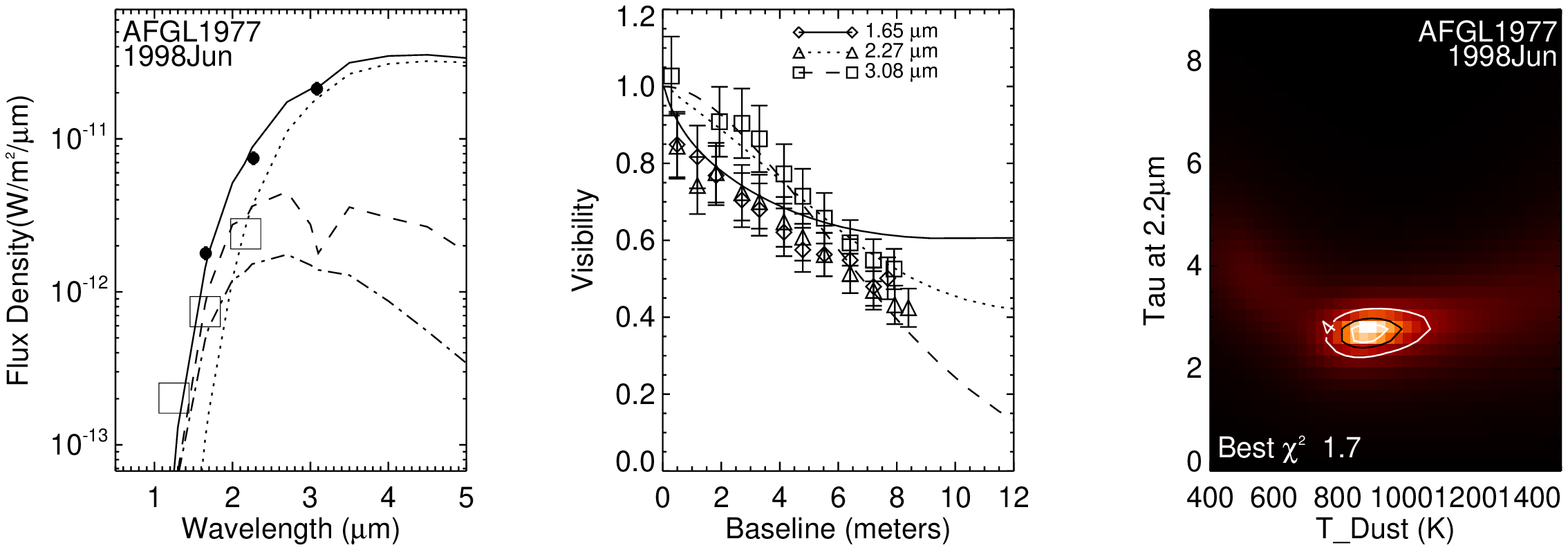}
\includegraphics[angle=90,width=3in]{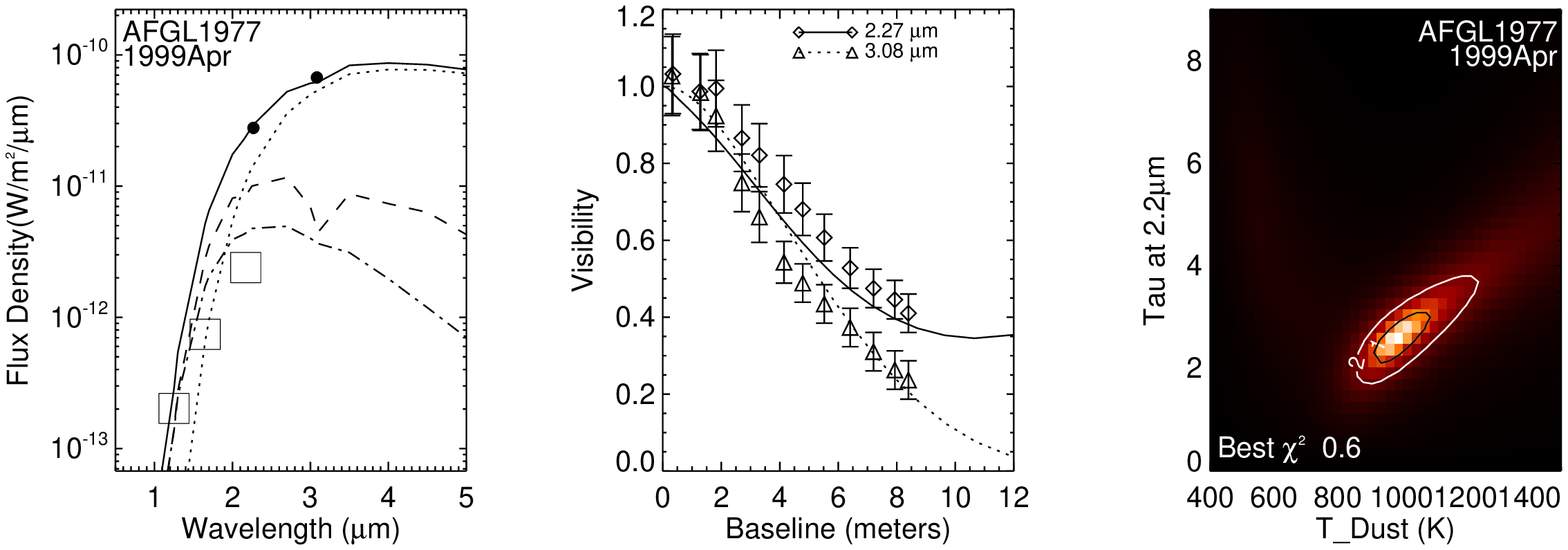}
\caption{ Best fit plots for AFGL1977.  See Fig.1 caption.}
\end{figure}

\begin{figure}
\centering
\includegraphics[angle=90,width=3in]{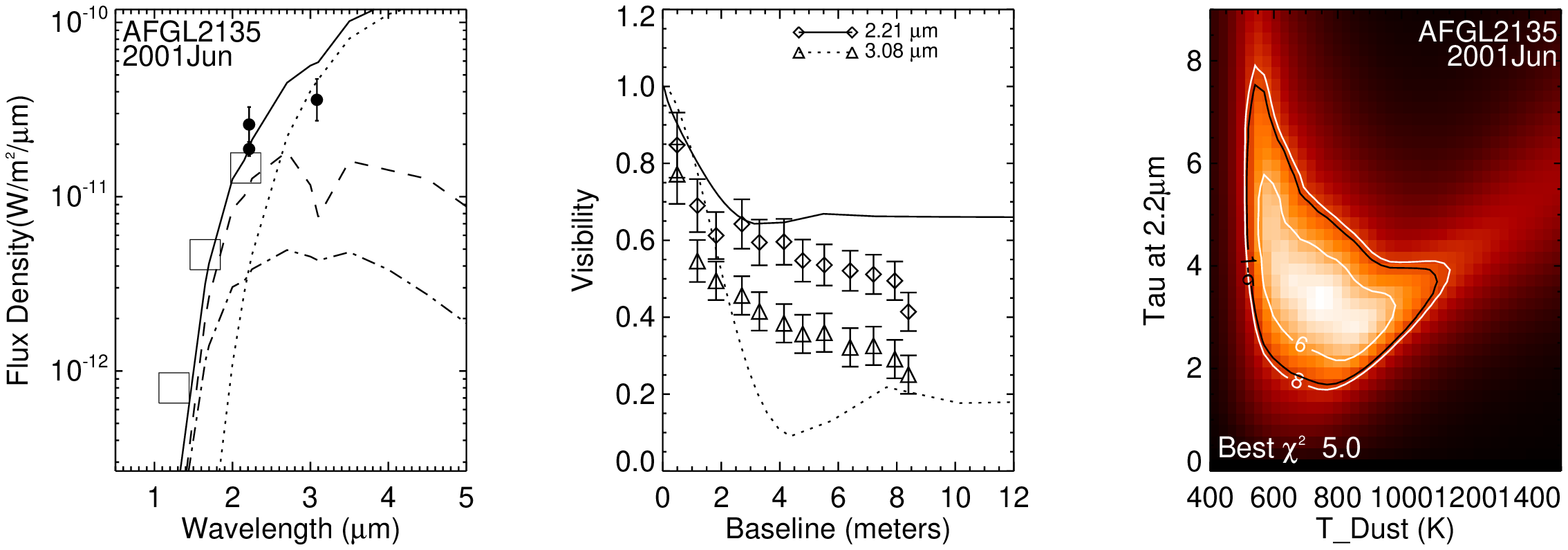}
\caption{ Best fit plots for AFGL2135.  See Fig.1 caption.}
\end{figure}

\begin{figure}
\centering
\includegraphics[angle=90,width=3in]{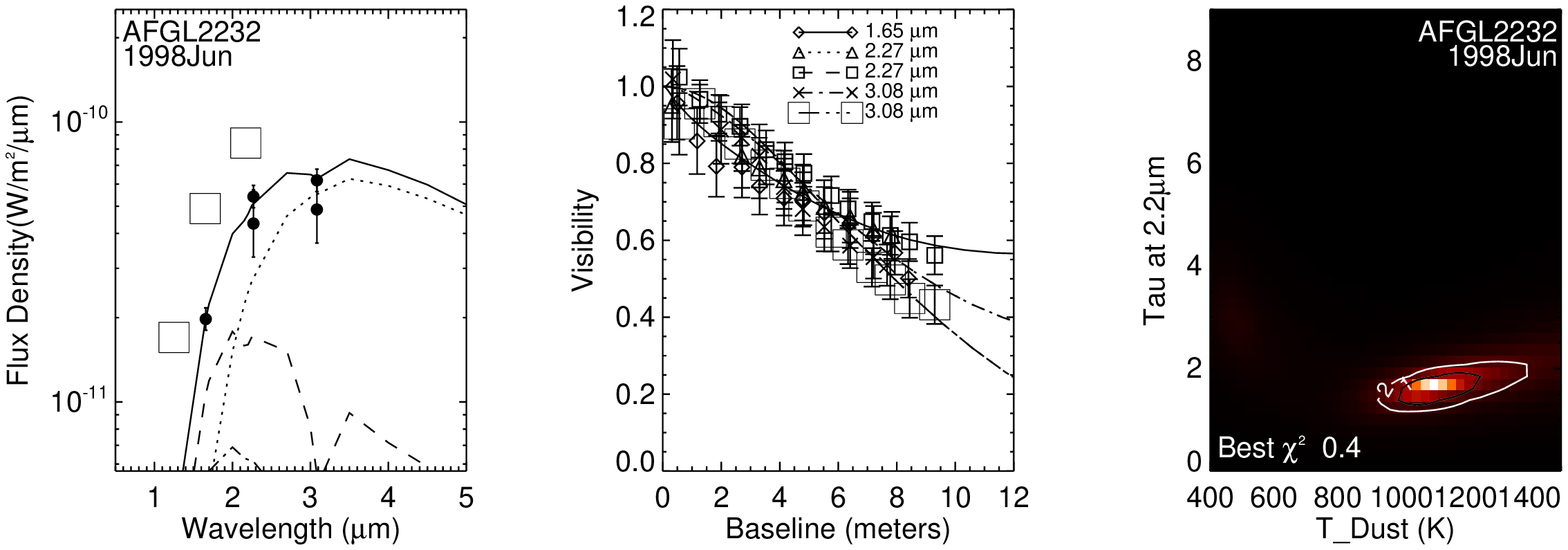}
\includegraphics[angle=90,width=3in]{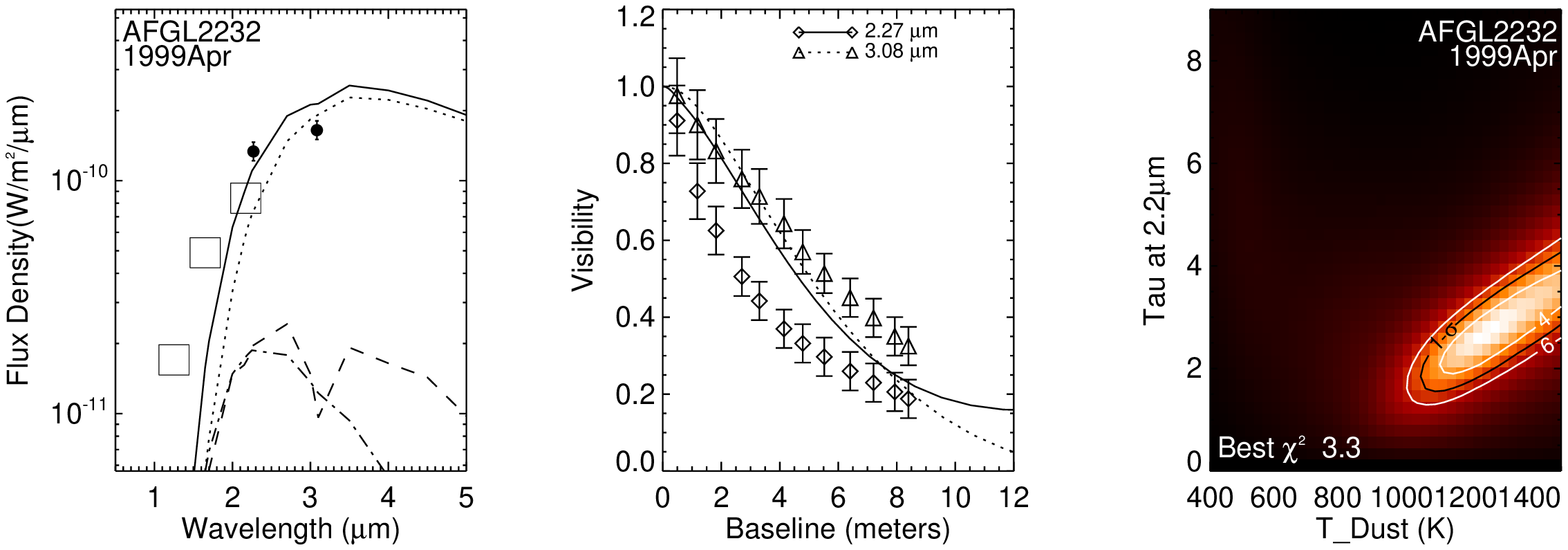}
\caption{ Best fit plots for AFGL2232.  See Fig.1 caption.}
\end{figure}

\begin{figure}
\centering
\includegraphics[angle=90,width=3in]{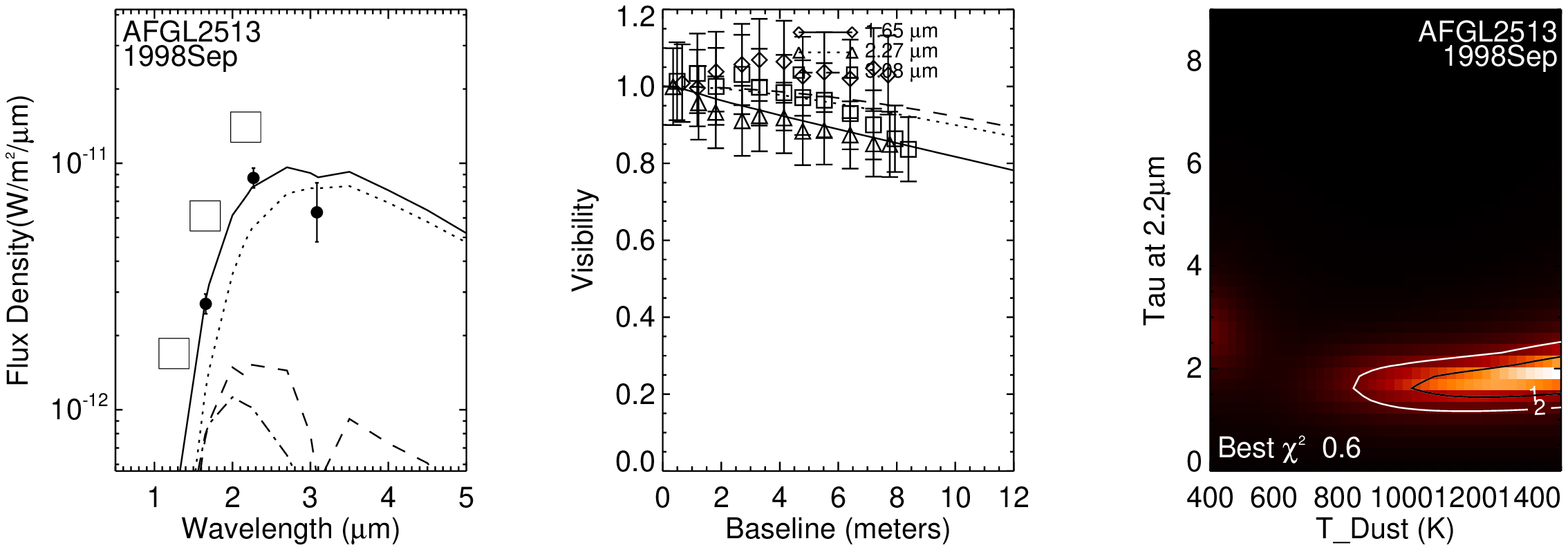}
\includegraphics[angle=90,width=3in]{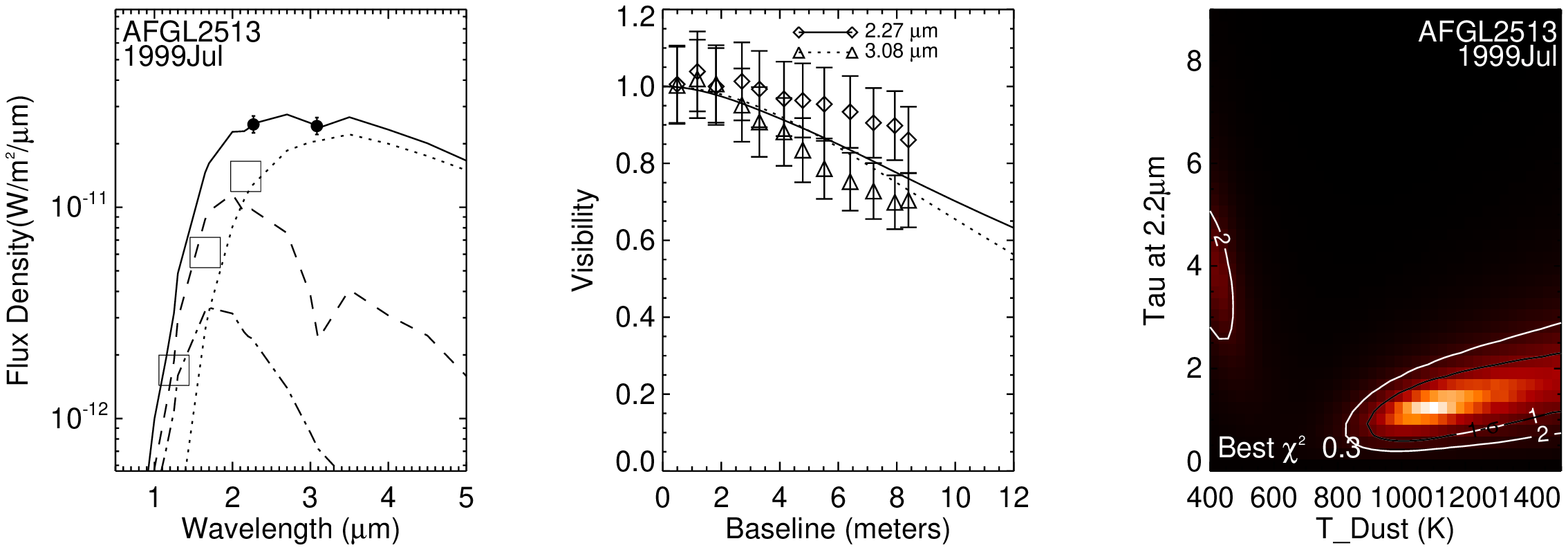}
\caption{ Best fit plots for AFGL2513.  See Fig.1 caption.}
\end{figure}

\begin{figure}
\centering
\includegraphics[angle=90,width=3in]{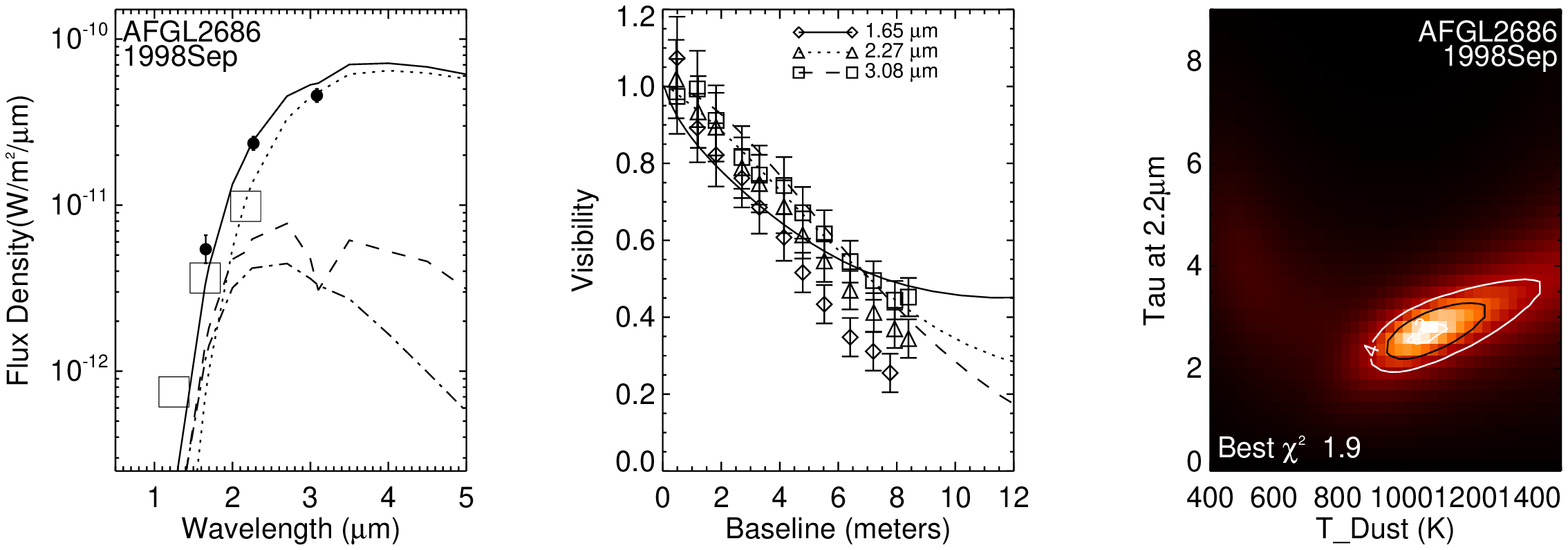}
\includegraphics[angle=90,width=3in]{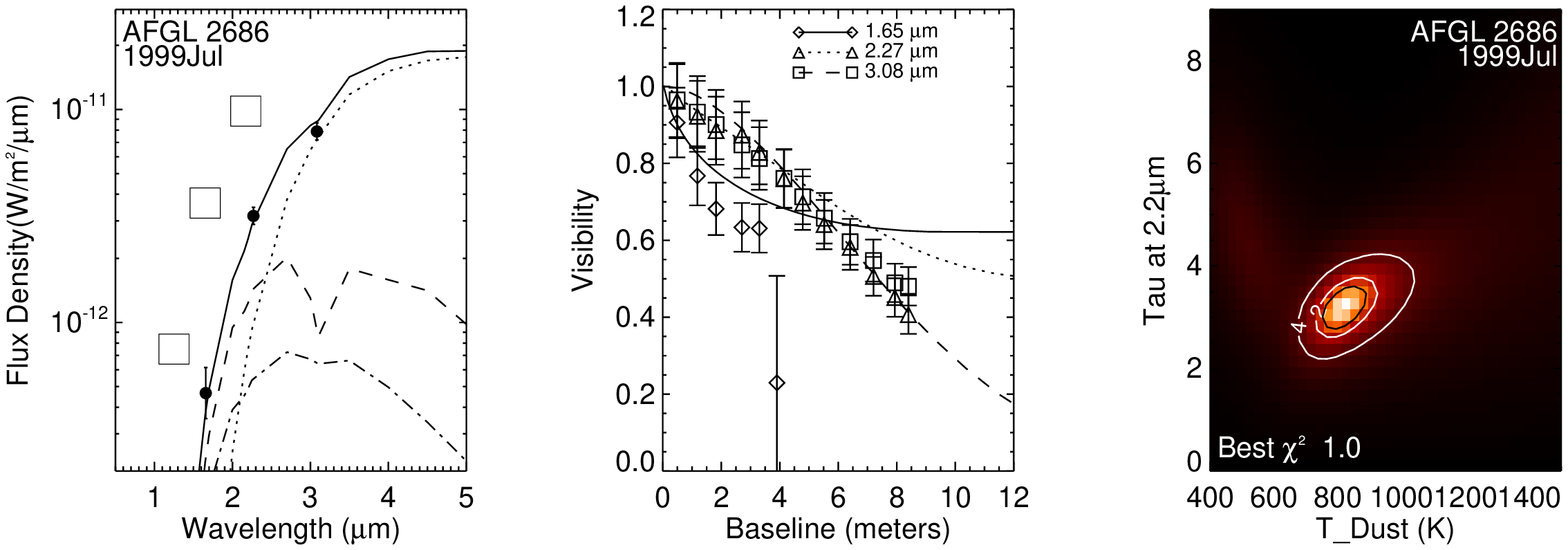}
\caption{ Best fit plots for AFGL2686.  See Fig.1 caption.}
\end{figure}

\clearpage

\begin{figure}
\centering
\includegraphics[angle=90,width=3in]{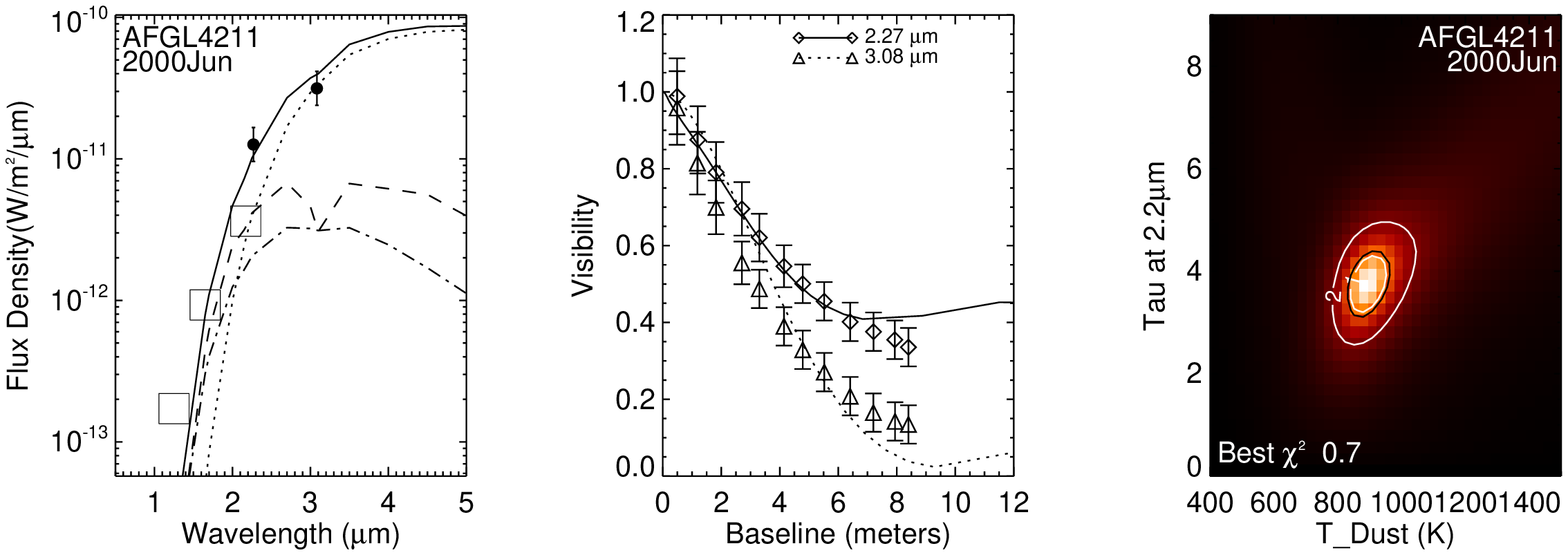}
\includegraphics[angle=90,width=3in]{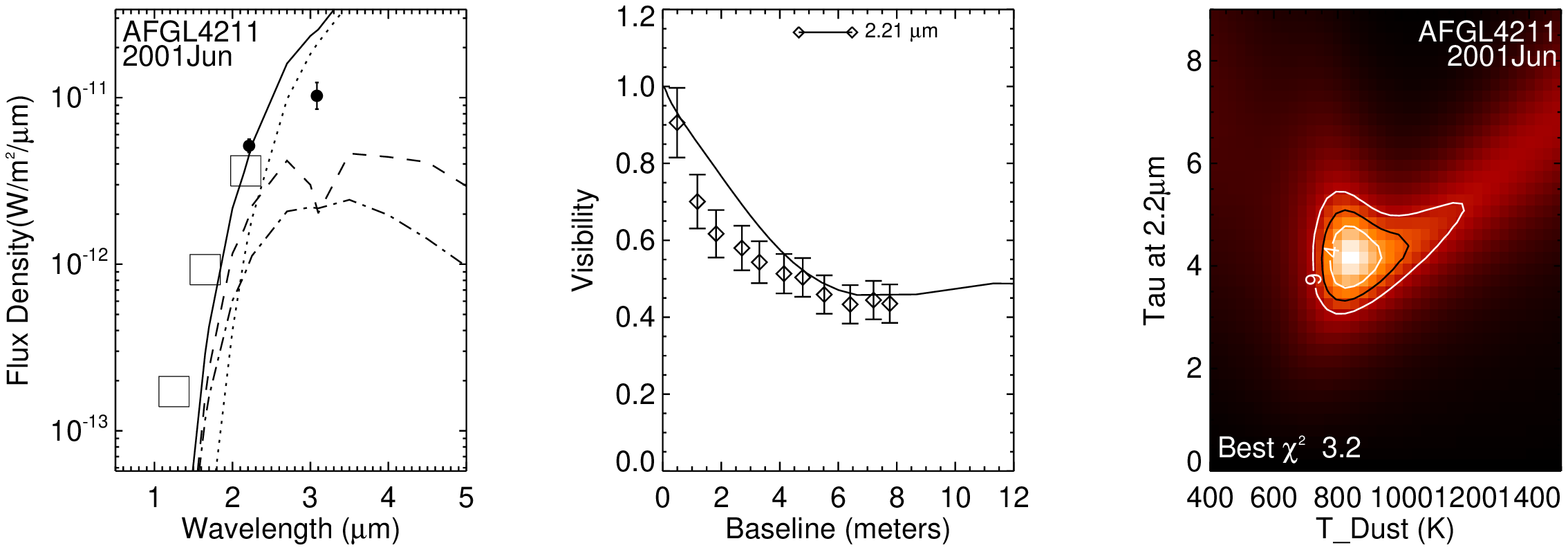}
\caption{ Best fit plots for AFGL4211.  See Fig.1 caption.}
\end{figure}


\begin{figure}
\centering
\includegraphics[angle=90,width=3in]{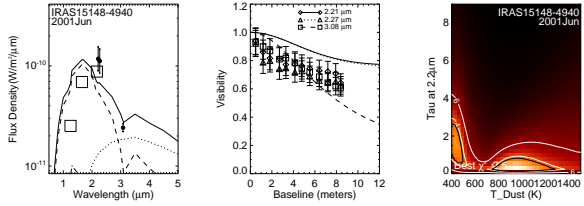}
\caption{ Best fit plots for IRAS15148-4940.  See Fig.1 caption.  We chose the lower-right region as
the best fit region because in all other cases of multiple good fitting regions the one at low tau and high dust temperature
was the consistent region.}
\end{figure}

\begin{figure}
\centering
\includegraphics[angle=90,width=3in]{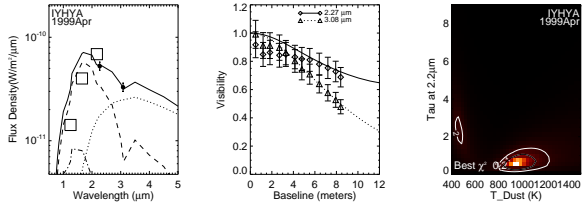}
\caption{ Best fit plots for IY~Hya.  See Fig.1 caption.}
\end{figure}

\begin{figure}
\centering
\includegraphics[angle=90,width=3in]{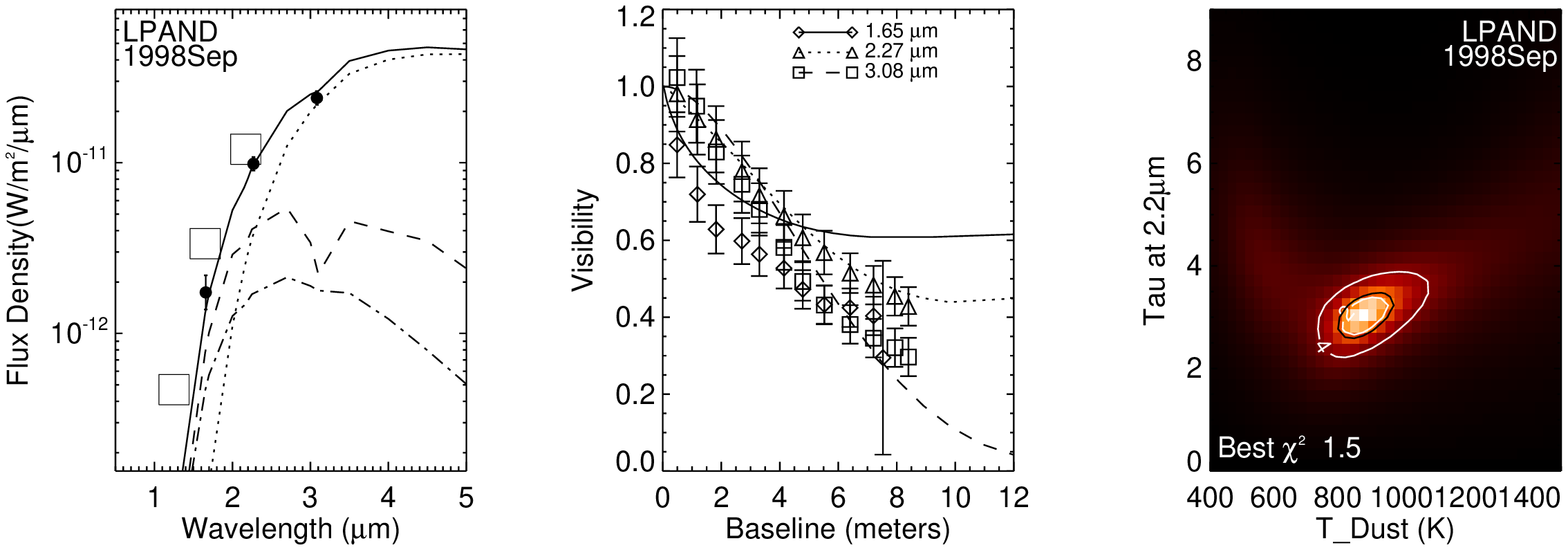}
\includegraphics[angle=90,width=3in]{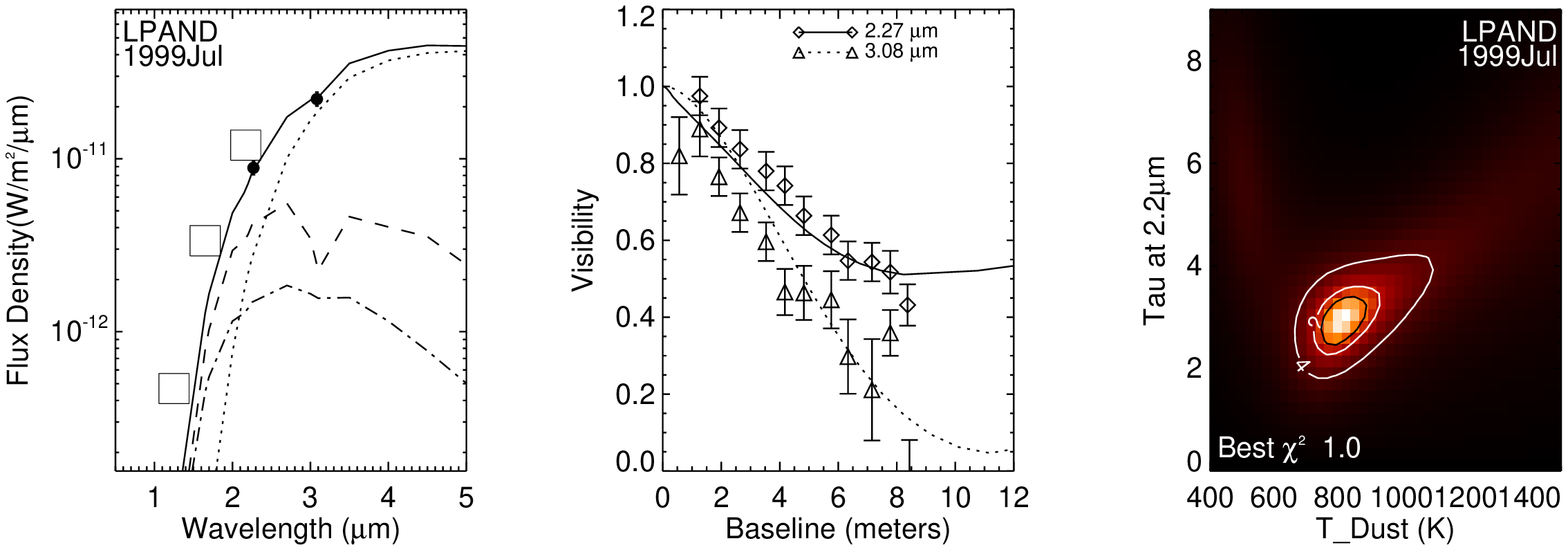}
\includegraphics[angle=90,width=3in]{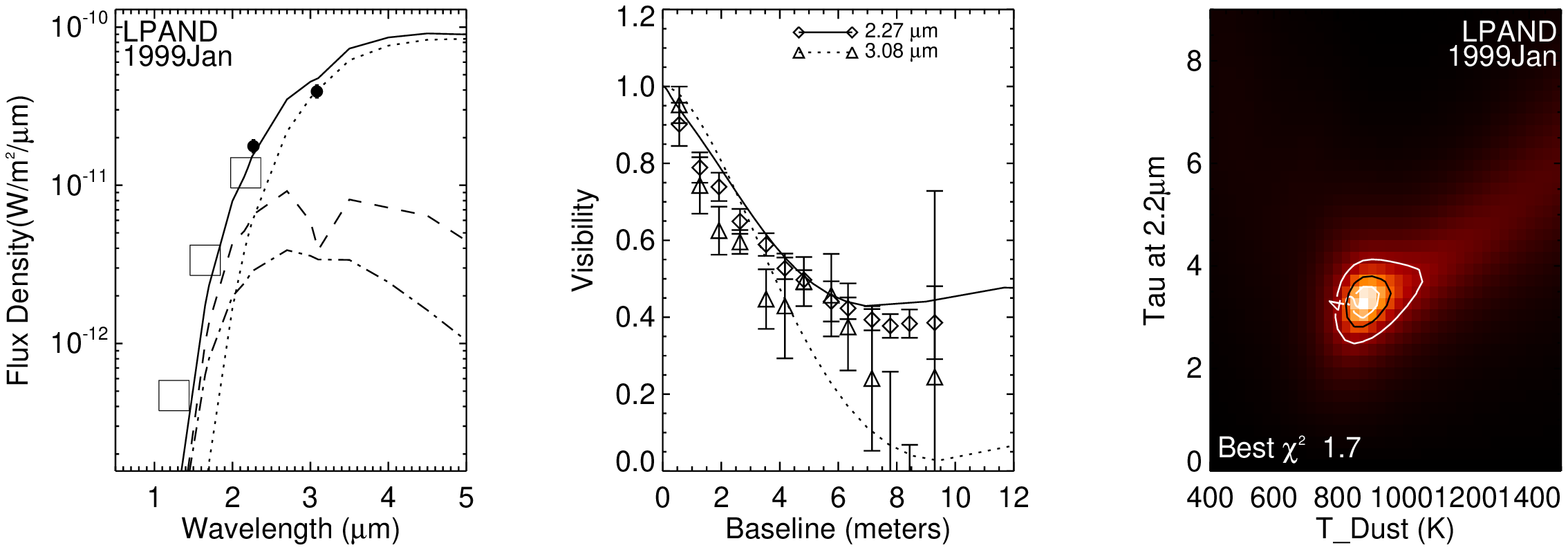}
\caption{ Best fit plots for LP~And.  See Fig.1 caption.}
\end{figure}

\begin{figure}
\centering
\includegraphics[angle=90,width=3in]{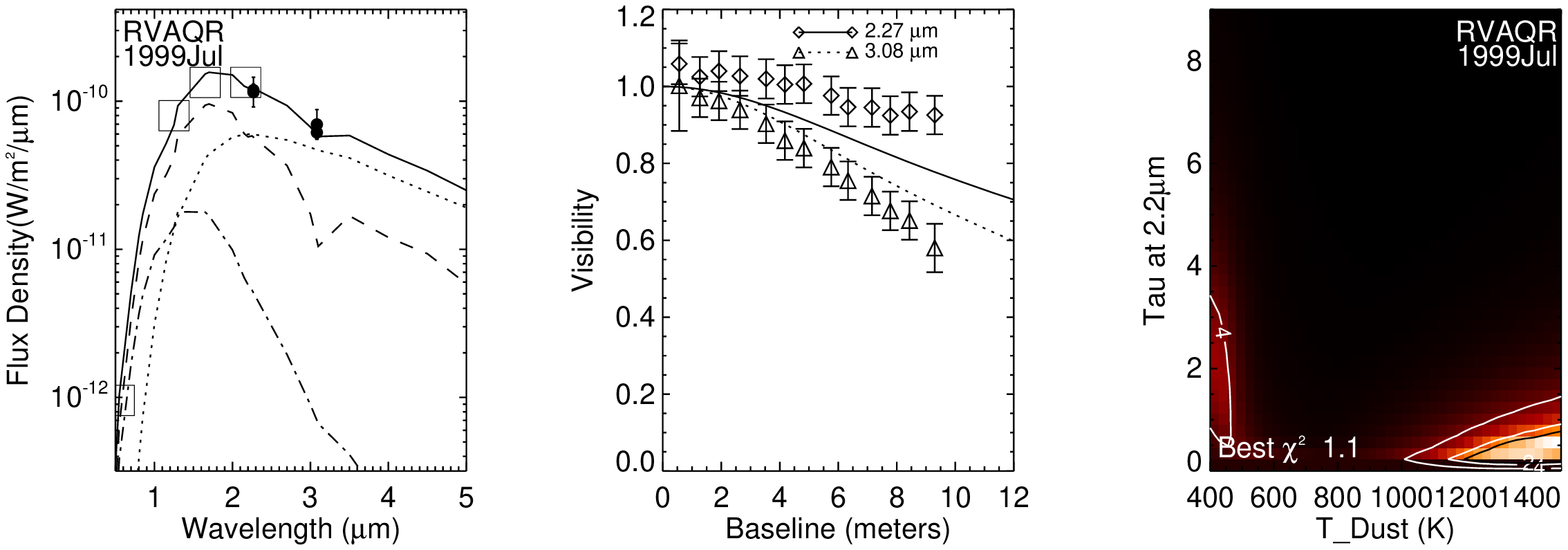}
\includegraphics[angle=90,width=3in]{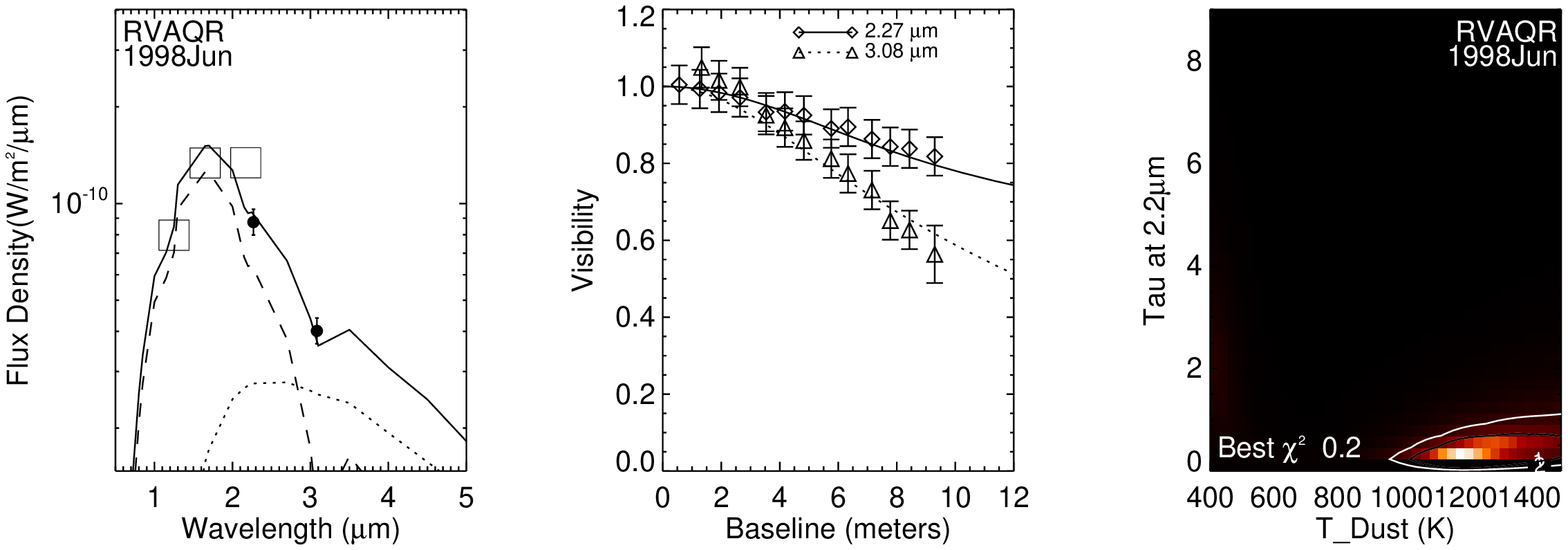}
\caption{ Best fit plots for RV~Aqr.  See Fig.1 caption.}
\end{figure}

\begin{figure}
\centering
\includegraphics[angle=90,width=3in]{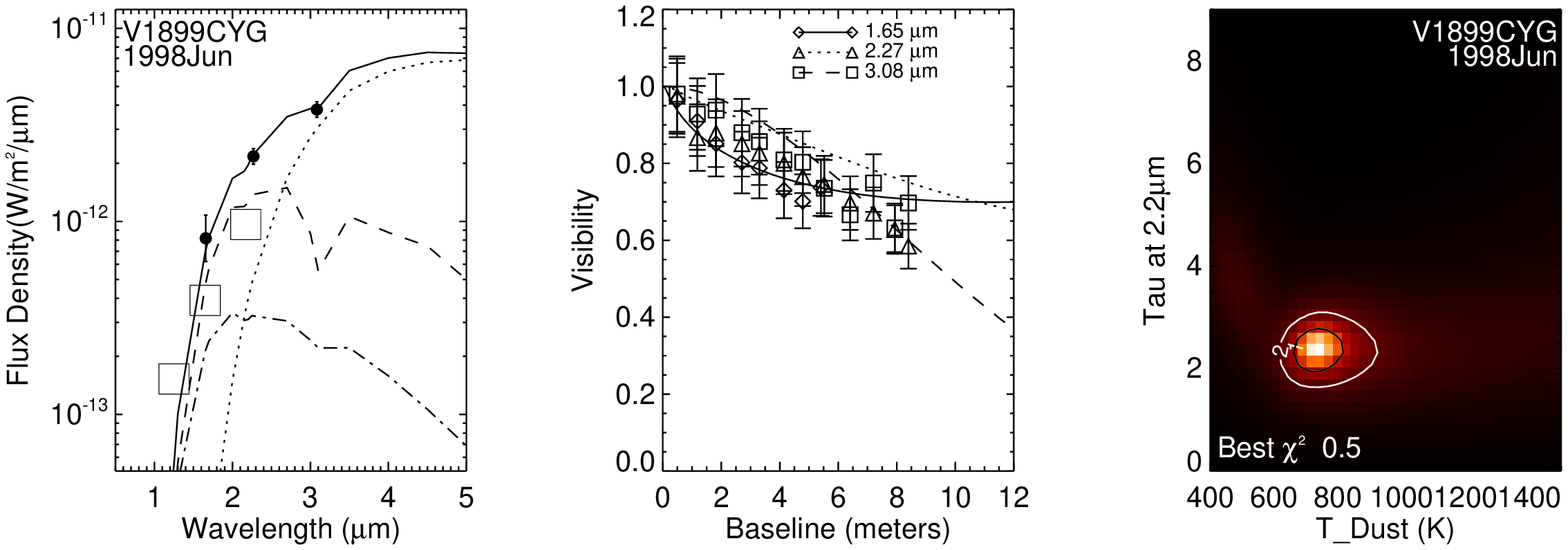}
\includegraphics[angle=90,width=3in]{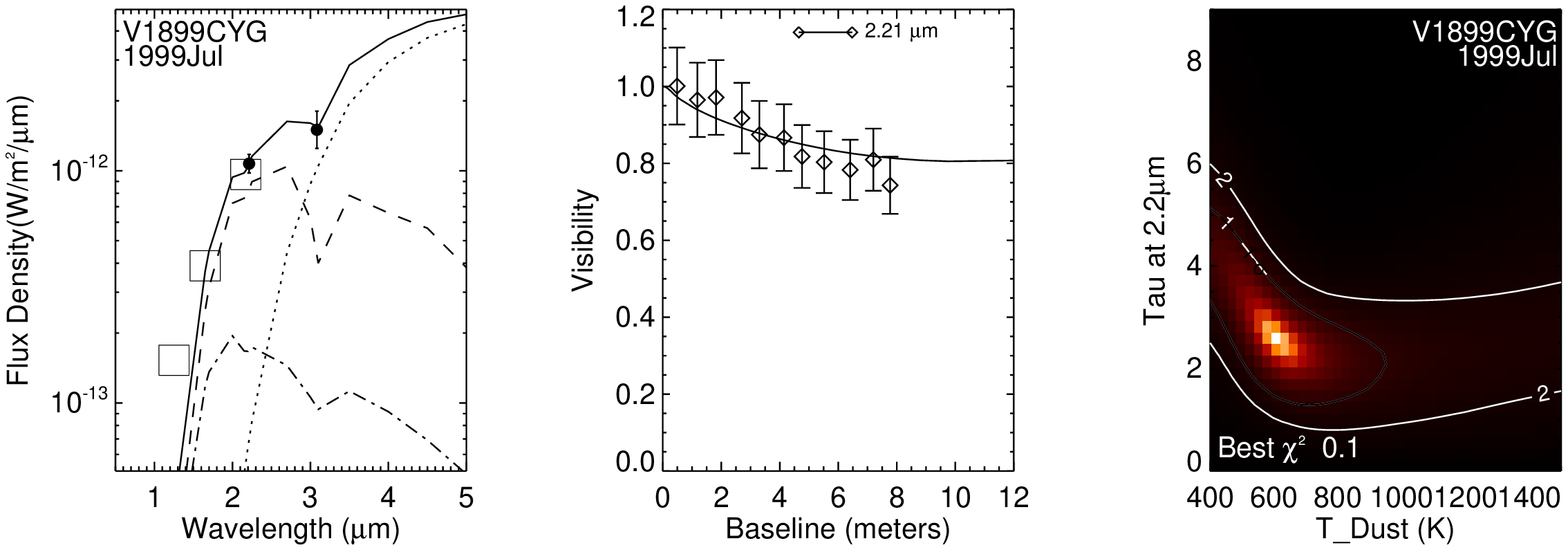}
\caption{ Best fit plots for v1899~Cyg.  See Fig.1 caption.}
\end{figure}

\begin{figure}
\centering
\includegraphics[angle=90,width=3in]{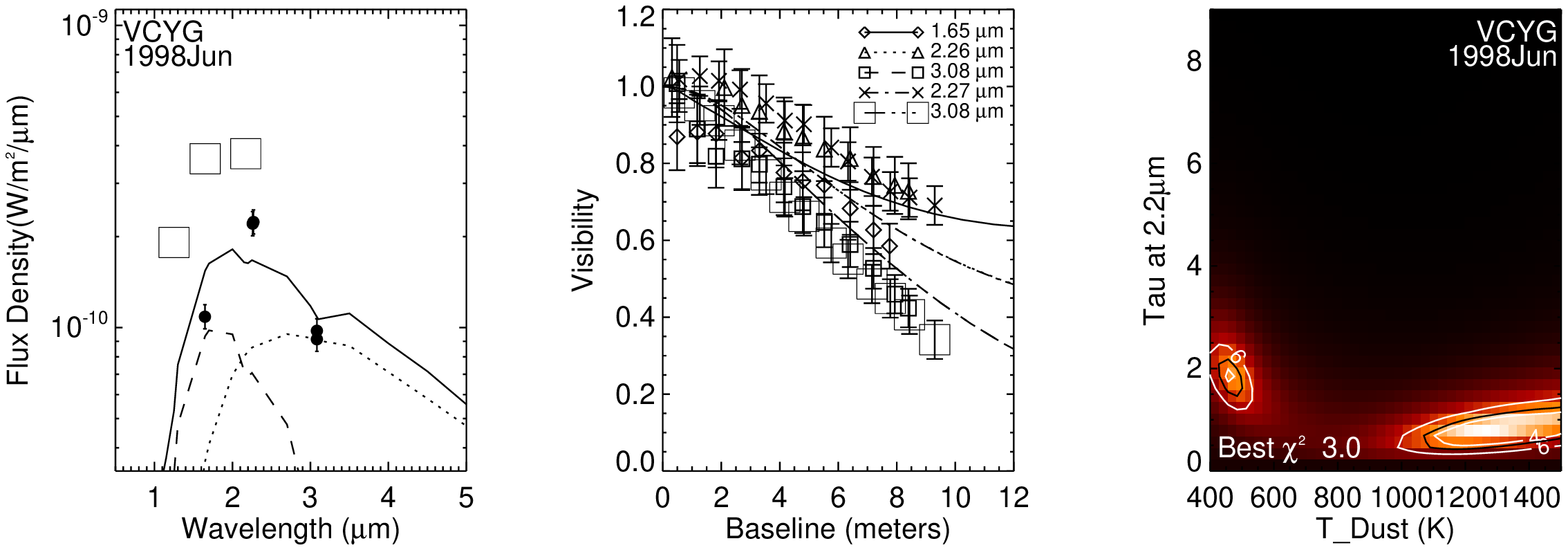}
\includegraphics[angle=90,width=3in]{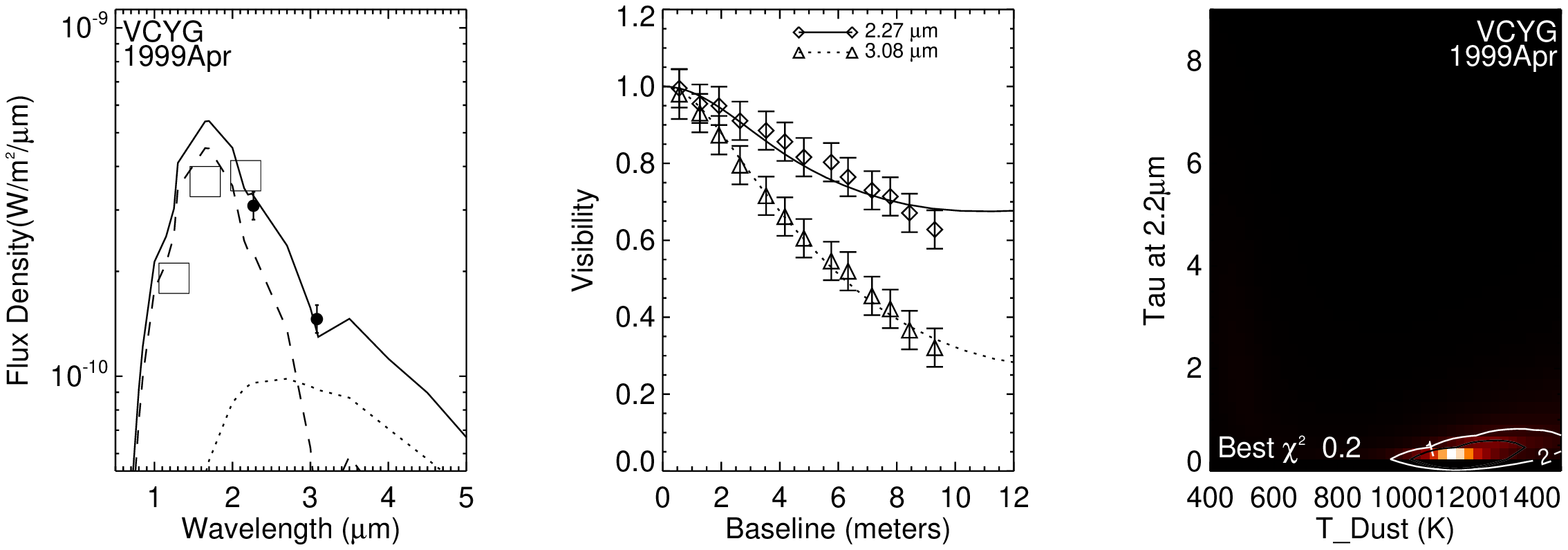}
\includegraphics[angle=90,width=3in]{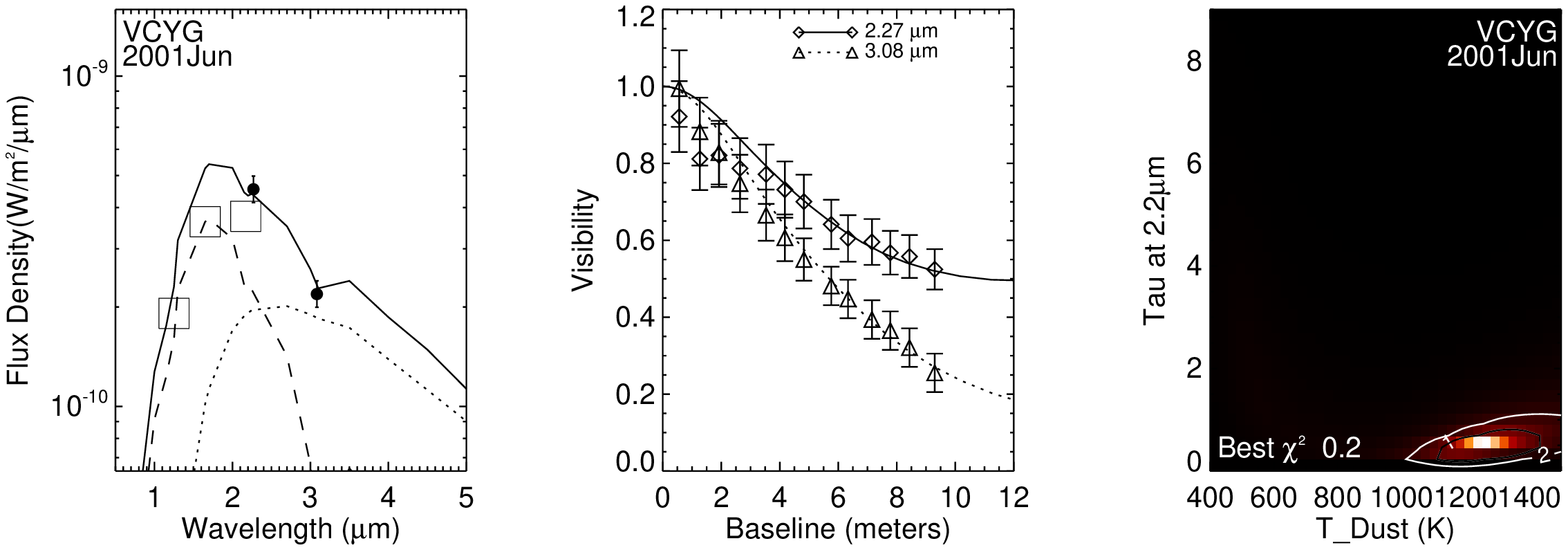}
\caption{ Best fit plots for V~Cyg.  See Fig.1 caption.  For the 1998 Jun. epoch the best fitting region was chosen to be the lower-right region because it is consistent with the other epochs.}
\end{figure}

\begin{figure}
\centering
\includegraphics[angle=90,width=3in]{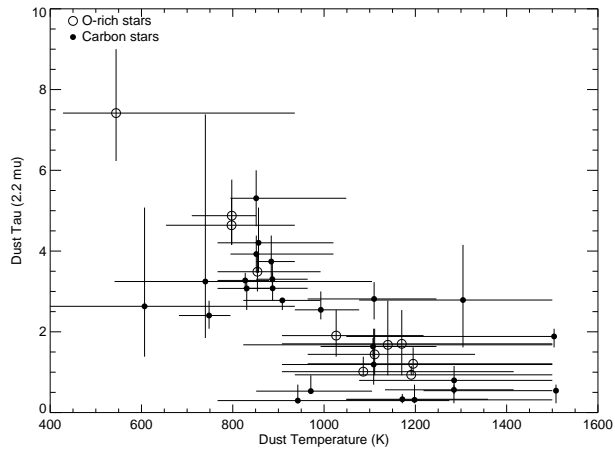}
\caption{ A plot of best fit $\tau_{2.2\mu\,m}$ versus the temperature at the inner edge of dust shell T$_{\rm dust}$. Open symbols are used for
oxygen-rich dust shells and closed symbols are used for carbon-rich dust shells.}
\end{figure}                        

\end{document}